\documentclass[pra,reprint,superscriptaddress]{revtex4-2}
\usepackage{epsfig}
\usepackage{graphicx}
\usepackage{amsmath}
\usepackage{amssymb}
\usepackage{mathrsfs}
\usepackage{dcolumn}
\usepackage{amsfonts}
\usepackage{ulem}
\newcommand{\Tr}{\mathrm{Tr}}
\newcommand{\ket}[1]{|#1\rangle}
\newcommand{\bra}[1]{\langle#1|}

\begin{document}
\title{Memory effects in a dynamical decoupling process}
\author{S. C. Hou}
\email{ housc@dlmu.edu.cn}
\affiliation{School of Science, Dalian Maritime University, Dalian 116026, China}
\author{X. Y. Zhang}
\affiliation{School of Science, Dalian Maritime University, Dalian 116026, China}
\author{Si-wen Li}
\affiliation{School of Science, Dalian Maritime University, Dalian 116026, China}
\author{X. X. Yi} 
\email{ yixx@nenu.edu.cn}
\affiliation{Center for Quantum Sciences and School of Physics, Northeast Normal University, Changchun 130024, China}
    
\begin{abstract}
We establish a simple quantitative relationship between the environmental memory effects and the 
characteristics in a dynamical decoupling process. In contrast to previous works, our measures of
non-Markovianity are tailored and extended to evaluate the strength of memory effects in dynamical 
decoupling.  We find that if each kick commutes with the dynamical map of the uncontrolled system, 
then the change of the final dynamical map or the final state brought by the control (called the 
``effect of control") is upper (lower) bounded by the summation (difference) of the strengths 
of memory effects with and without control. We propose sufficient conditions for the 
commutation relation for parity kicks and illustrate our finding with a dissipative quantum Rabi
model by numerical simulations where one or many cycles of parity kicks are implemented on the qubit. 
Besides, the results show that under certain conditions, the effect of control or the increase of 
performance by the control may be simply proportional to the strength of memory effects with or 
without control. 
\end{abstract}
\date{\today}
\maketitle

\section{Introduction}

Dynamical decoupling \cite{Viola1998,Viola,Vitali1999,Zanardi,Rego,Souza,Lidar2014,Ezzell2023}, 
introduced in the late 2000s, is a powerful open-loop control technique where periodic fast unitary operations
(called ``kicks") are implemented on a quantum system to protect it against environmental noise, such as
decoherence and dissipation. It originates from the ``spin-echo" or ``refocusing" phenomena in nuclear magnetic 
resonance \cite{Hahn}.  Since its introduction, the technique has been extensively studied theoretically and 
experimentally to protect quantum states or quantum gates \cite{Viola2003,Khodjasteh,Uhrig,West,Souza2011,Du,Biercuk,Lange,Zhang2015}.

It is well known that the non-Markovian character of the environment (memory effects) is crucial for the success 
of dynamical decoupling since the period of the kicks has to be small compared with the fastest time scale of the 
interactions with the environment \cite{Viola1998,Viola,Vitali1999,Zanardi}. From  an informational
 perspective, in the time scale of the period of the kicks, the information lost from the system has to be preserved in the environment 
and exploited to influence the system's future dynamics. Therefore, the system evolution under control depends on its 
history which is the very essence of the non-Markovian character. In recent years, the interplay between dynamical 
decoupling and the environmental memory effects has received much attention \cite{Addis,Arenz,Szczygielski,Gough,Dong2018,Khurana,Berk,Burgarth,Burgarth2021}.
On one hand, the investigation of this issue is beneficial for the advance of quantum information processing.  
On the other hand, it sheds light on the concept of quantum non-Markovianity \cite{Rivas2014,Breuer2016,Vega2017,Li2018} 
which is of fundamental interest. 

Particularly, the quantitative relations between the characteristics of dynamical decoupling and strength of 
environmental memory effects (non-Markovianity) have been studied in several works, yet not so extensively.
For example, the efficiency of dynamical decoupling is compared with the Breuer-Laine-Pilo measure of non-Markovianity 
\cite{Breuer2009} in uncontrolled or controlled dynamics in a dephsaing model in Refs.\ \cite{Addis} and \cite{Khurana}. 
This measure \cite{Breuer2009} is defined as the total increments of distinguishability in time evolutions maximized over 
all the possible pair of initial states and associated with the information backflow. However, their results indicate 
that the dependence of the control efficiency on the non-Markovianity is not quite straightforward \cite{Addis,Khurana} 
(neither proportional nor monotonic). Recently, using quantum resource theories \cite{Chitambar,Berk2021}, the authors 
in Ref.\ \cite{Berk} show that quantum non-Markovianity, interpreted as the multitime correlations, is a resource for 
dynamical decoupling. The theory is formulated in the framework of process tensor whose characterization scales 
exponentially with the number of kicks. In their work, the kicks are assumed to be free transformations that do 
not increase the non-Markovianity and the success of dynamical decoupling is quantified by the total temporal correlations in the process 
tensor. Before and after a coarse-graining process to a smaller number of interventions \cite{Berk}, it is seen that the 
increase of the total temporal correlations in the process tensor is accompanied by the consumption of non-Markovian 
correlations. 

Given the conceptual and computational complexities in Ref.\ \cite{Berk}, the aim of this paper is to establish a 
simple relation that quantitatively links the dynamical decoupling characteristics and the strength of memory effects. 
To accomplish this, we adopt and extend the measures of non-Markovianity proposed in our previous works 
\cite{Hou2015,Hou2024} to quantify the strength of memory effects in a dynamical decoupling process (or that without control).
Originally, our non-Markovianity measures are defined as the maximal difference of completely positive 
dynamical maps $T(t_2,t_0)$ and $T(t_2,t_1)T(t_1,t_0)$ \cite{Hou2015} (or those acting on a system initial state \cite{Hou2024})
by optimizing over $t_1$ and $t_2$ ($t_0$ is fixed). In this work, we fix $t_1$ and $t_2$ to quantify the memory effects in a 
dynamical decoupling process with two kicks at $t_1$ and $t_2$ and further extend our measures to their multitime
forms for more kicks. Compared with the measures used in Refs.\ \cite{Addis,Khurana} or \cite{Berk}, the measures 
in this work have the following merits. First, they can be directly interpreted as the 
past-future dependence, i.e., the dependence of the system's future dynamics on its history, 
which is the essence of non-Markovianity. Second, the calculation of memory effects is 
computationally efficient since the description of the dynamical maps \cite{Hou2015} or 
quantum states \cite{Hou2024} in our measures does not scale with the number of kicks and requires 
no optimizations over the initial states \cite{Breuer2009} or time instants \cite{Hou2015,Hou2024}. 
Last, our measures reflect the quality of the Born-Markov approximation for a quantum process \cite{Hou2015,Hou2024}.

Intuitively, the dynamical decoupling of a near-Markovian quantum process leads to a bad performance (e.g., low fidelity 
between the final state and the target state.) and very weak memory effects, meanwhile, a perfectly decoupled dynamics 
may become effectively unitary and thus Markovian, again leading to very weak memory effects.  
Thus there may not exist a monotonic relation between the dynamical decoupling 
performance and the strength of memory effects in the controlled process. Instead, we focus on 
the change of the system's final dynamical map (final state) brought by dynamical decoupling, which we called the ``effect of control", 
and the strengths of memory effects in both the controlled and uncontrolled processes. Our main finding is that if each kick commutes 
with the dynamical map of the uncontrolled system, then the effect of control is upper (lower) bounded by the summation (difference) of 
the strengths of memory effects in the controlled and uncontrolled processes. That is, the non-Markovianity of the 
controlled and uncontrolled process together contribute to the change of the final dynamical map (final state) in general.
We propose a class of models where a parity kick \cite{Vitali1999} commutes with the dynamical map without control. 
After that, we illustrate our finding by numerically simulating a dissipative quantum 
Rabi model where parity kicks are applied to the qubit to eliminate the influence of its environment (an oscillator
in a Markovian thermal reservoir). Results for one and many cycles of parity kicks in different regimes coincide with our theory. 
Moreover, we find that, in many cases, the effect of control may be approximately equal to the increase of 
performance brought by the control and proportional to the strength of memory effects in the controlled or 
uncontrolled process.

The paper is organized as follows: In Sec.\ \textrm{II} we briefly introduce the 
theory of parity kicks and the standard dynamical decoupling. In Sec.\ \textrm{III},
our non-Markovianity measures are reviewed, tailored and extended to evaluate the memory effects 
in dynamical decoupling. Next, these measures are applied to a dynamical decoupling process 
with two or more kicks in Sec.\ \textrm{IV} to establish quantitative connections 
between the effect of control and the strengths of memory effects. In Sec.\textrm{V}, 
we propose sufficient conditions for the commutation between a parity kick and 
the uncontrolled dynamical map and illustrate our finding with a dissipative quantum Rabi model 
in different regimes. At last, we summarize our work in Sec.\ \textrm{VI}.   


\section{Dynamical Decoupling}
We consider an open quantum system described by the time-independent system-environment Hamiltonian
\begin{eqnarray}
H_0 &=& H_S+H_E+H_{S\!E}
\label{Eqn:Hmt}
\end{eqnarray}
where $H_S$ is the Hamiltonian of the interested system and $H_E$ is that of its environment. 
The interaction of the system and environment is described by 
\begin{eqnarray}
H_{S\!E}=\sum_n S_n\otimes E_n
\label{Eqn:InterHmt}
\end{eqnarray}
which is unwanted. The dynamical decoupling is implemented by a periodic time-dependent control 
Hamiltonian $H_1(t)$ acting  on the system only.  Then the total Hamiltonian is given by  
\begin{eqnarray}
H_c(t) &=& H_0+H_1(t)
\label{Eqn:HmtControl}
\end{eqnarray}
and the system-environment evolution under control is determined by $U(t)=\mathcal{T}e^{-i\int_{0}^{t}H_c(\tau)d\tau}$  
where $\mathcal{T}$ is the time-ordering operator ($\hbar=1$ throughout the paper).

\subsection{Parity kicks}
We first introduce the decoupling scheme by parity kicks \cite{Vitali1999} and 
try to find  quantitative connections between the environmental memory effects 
and the decoupling process in one cycle of (two) parity kicks later in Sec.\ \ref{Sec:MemoDD} A. 
The reason is that the dynamical decoupling process by two kicks is relevant to how the system's 
history (before the first kick) influences its future (after the second kick).  
This is exactly the physical meaning of memory effects defined in Refs.\ \cite{Hou2015,Hou2024}.
In fact, our theory also applies to a general dynamical decoupling process with 
more than two kicks. In the latter case, the dynamical decoupling process is associated with 
the memory effects from the perspective of the environment as discussed in Sec.\ \ref{Sec:Measure} B.  

The control Hamiltonian for one cycle of parity kicks is given by 
\begin{eqnarray}
H_1(t)=\sum_{j=1}^{2}\delta(t-t_j)H_P,
\label{Eqn:CHParity}
\end{eqnarray}
which represents 2 instantaneous pulses at $t_j=j\tau$. 
Let the ``width" of the delta function be $\tau_0$. In the ideal 
case ($\tau_0\rightarrow0$),  $H_0$ is ignorable 
during a pulse so that the parity kick can be realized by $K_P=e^{-iH_P\tau_0}$ 
with properly chosen $H_P$ and $\tau_0$.
Note that a parity operator satisfies $K_P=K_P^{\dag}=K_P^{-1}$. It is known  that if the 
following conditions hold,
\begin{eqnarray}
K_PH_SK_P&=&H_S,
\label{Eqn:ParityKickConditon1}\\
K_PH_{S\!E}K_P&=&-H_{S\!E}, 
\label{Eqn:ParityKickConditon2}
\end{eqnarray}
the system can be protected from the influence of the environment at the end of 
the $2j$th kick \cite{Vitali1999}. The mechanism can be explained by the 
total time evolution in one cycle of two kicks:
\begin{eqnarray}
U(t_c)&=&K_Pe^{-iH_0\tau}K_Pe^{-iH_0\tau}\nonumber\\
 &=&e^{-iK_PH_0K_P\tau}e^{-iH_0\tau}\nonumber\\
 &=&e^{-i(H_S+H_E-H_{S\!E})\tau}e^{-i(H_S+H_E+H_{S\!E})\tau}
\label{Eqn:PKCycle}
\end{eqnarray}
where $t_c=2\tau$ is the cycle time. It is shown that the sign of $H_{S\!E}$ 
in the second half of the cycle is effectively changed by the parity kicks.
Thus the influence of the environment on the evolution is effectively time-reversed 
after the first kick. In the limit of continuous kicks ($t_c\rightarrow0$), 
there are \cite{Vitali1999}   
\begin{eqnarray}
U(t_c)=e^{-i(H_S+H_E)t_c}.
\label{Eqn:LimitPKCycle}
\end{eqnarray}
In practice, if the cycle time $t_c$ is sufficiently small compared with the 
time scale characterizing the unwanted interactions with the environment
(non-Markovian condition), the time reversal is expected to be effective and 
the influence of $H_{S\!E}$ can be eliminated. For particular models, e.g., the 
Jaynes-Cummings (JC) model where the single-mode field serves as the environment, 
it is seen that the parity kicks can perfectly eliminate the interaction with the 
environment with any cycle time if there is no dissipation in the single-mode field \cite{Morigi,Meunier}.  
The above facts imply that the environmental memory effects are important 
for dynamical decoupling since the system's information has to be well preserved 
in the time scale of the cycle time $t_c$.

\subsection{General case}
In general, a standard dynamical decoupling contains cycles of instantaneous kicks.  
Each kick implements a particular unitary transformation on the system to protect it 
against the environment noise. The time evolution of the system and environment 
in one cycle of kicks can be designed by 
\begin{eqnarray}
U(t_c)&=& K_{m}e^{-iH_0\tau}\cdots K_{2}e^{-iH_0\tau}K_{1}e^{-iH_0\tau}
\label{Eqn:GeneralDD}
\end{eqnarray}
where $t_c=m\tau$ is the cycle time and $K_j$ ($j=1,2,\cdots m$) represent $m$ instantaneous kicks.
The corresponding control Hamiltonian is given by  
\begin{eqnarray}
H_1(t)=\sum_{j=1}^{m}\delta(t-t_j)H_{K_j}.
\label{Eqn:CHDD}
\end{eqnarray}
Similarly, an ideal kick can be realized by $K_j=e^{-iH_{K_j}\tau_0}$ in the limit $\tau_0\rightarrow0$.

In general, the dynamical decoupling pulses can be constructed based on symmetrization
 \cite{Viola,Zanardi,Lidar2014,Ezzell2023,Rego}. Let $\mathcal{G}=\{g_j\}$ ($j=1,2,\cdots,n_g$) 
be a finite group of unitary transformations acting on the system where $g_1=I$ is the identity 
transformation and $n_g=|\mathcal{G}|$ is the number of the group elements [in Eq.\ (\ref{Eqn:GeneralDD}), $m=n_g$]. 
Then one cycle of dynamical decoupling can be constructed by
\begin{eqnarray}
U(t_c)&=&  g^{\dag}_{n_g}e^{-iH_0\tau}g_{n_g}\cdots g^{\dag}_2e^{-iH_0\tau}g_2g^{\dag}_1e^{-iH_0\tau}g_1\nonumber\\
      &=&  e^{-i\tau g^{\dag}_{n_g}H_0g_{n_g}}\cdots e^{-i\tau g^{\dag}_2H_0g_2}e^{-i\tau g^{\dag}_1H_0g_1}\nonumber\\
      &=&e^{-i\tau \sum_{j=1}^{n_g}g^{\dag}_jH_0g_j}+\mathcal{O}(t_c^2)\nonumber\\
      &=&e^{-i t_c \bar{H}_0}+\mathcal{O}(t_c^2).
\label{Eqn:GeneralDDG}
\end{eqnarray}
Here $\bar{H}_0=\frac{1}{n_g}\sum_{j=1}^{n_g}g^{\dag}_jH_0g_j$ is the average Hamiltonian \cite{Viola} 
and the cycle time are $t_c=n_g\tau$. With properly designed group $\mathcal{G}$, the average Hamiltonian becomes harmless. 
For example, if $\bar{H}_0=H_S+H_E$ as in Eq.\ (\ref{Eqn:LimitPKCycle}), then the interaction $H_{S\!E}$ 
can be effectively eliminated at the end of the cycle when $t_c$ is small. 

According to Eq.\ (\ref{Eqn:GeneralDDG}), the kicks in Eq.\ (\ref{Eqn:GeneralDD}) are  
\begin{eqnarray}
K_{j}=\left \{
   \begin{array}{rcl}
  g_{j+1}g^{\dag}_{j}& &(j=1,2,\cdots n_g-1) \\
   g^{\dag}_{n_g}  & &(j=n_g) \\
 \end{array}
\right .
\label{Eqn:Groupdesign}
\end{eqnarray}
with $n_g=m$. Using the above construction,  Eq.\ (\ref{Eqn:PKCycle}) can be understood 
by $K_1=g_2g_1^{\dag}=K_P$ and $K_2=g^{\dag}_2=K_P$ where $g_1=I$ and $g_2=K_P$ are the 
elements of a group of order 2. Thus the parity kick scheme can be cast into 
the framework of dynamical decoupling. Using Eq.\ (\ref{Eqn:Groupdesign}), it is easy 
to see that the condition  
\begin{eqnarray}
K_{n_g}K_{n_g-1}\cdots K_1=g^{\dag}_{n_g}g_{n_g}g^{\dag}_{n_g-1}\cdots g_2g^{\dag}_1 =I
\label{Eqn:condition}
\end{eqnarray}
holds since $g^{\dag}_j=g^{-1}_j$ and $g_1=I$. Eq.\ (\ref{Eqn:condition}) demonstrates  
that the successive actions of the kicks leave the system invariant.


\section{Measure of memory effects for dynamical decoupling}
\label{Sec:Measure}
In this section, we review the measures and physical meaning of the non-Markovianity (interpreted as memory effects)
proposed in our previous works \cite{Hou2015,Hou2024} and propose a strategy to quantify the strength of memory 
effects in a dynamical decoupling process.   

\subsection{Measure of memory effects}
\label{Sec:MeasureA}
The measure of non-Markovianity defined in Ref.\ \cite{Hou2015} deals with a quantum process 
governed by a system-environment Hamiltonian (possibly time-dependent) $H(t)=H_S(t)+H_E(t)+H_{S\!E}(t)$
 like Eq. (\ref{Eqn:Hmt}) or Eq. (\ref{Eqn:HmtControl}), an arbitrary system initial state 
$\rho_S(t_I)$, a fixed initial state of the environment $\rho_E(t_I)$, and the condition  
\begin{eqnarray}
\rho_{S\!E}(t_I)=\rho_{S}(t_I)\otimes\rho_E(t_I),
\label{Eqn:InitialCondition}
\end{eqnarray}
where $t_I$ is an arbitrary initial time. In general, the initial state of the environment is given by \cite{Li2018}
\begin{eqnarray}
\rho_E(t_I)=\mathcal{T}e^{-i\int_{0}^{t_I}H_E(t')dt'}\rho_E(0).
\label{Eqn:EnvInitialState}
\end{eqnarray}
We define the universal dynamical map (UDM) \cite{Rivas} $T(t_b,t_a)$ as the dynamical map that transfers any 
$\rho_S(t_a)$ to $\rho_S(t_b)$ ($t_a\leqslant t_b$) \cite{Hou2015},  i.e.,
 \begin{eqnarray}
  \rho_S(t_b)&=&T(t_b,t_a)\rho_S(t_a)  \nonumber\\
  &=&\Tr_E[U(t_b,t_a)\rho_S(t_a)\otimes\rho_E(t_a)U(t_b,t_a)^{\dag}].
\label{Eqn:T}
\end{eqnarray}
where $t_a=t_I$ and $U(t_b,t_a)=\mathcal{T}e^{-i\int_{t_a}^{t_b}H(\tau)d\tau}$. Let $t_0\leqslant t_1\leqslant t_2$,
then the Markovian divisibility condition can be expresssed as \cite{Hou2015,Rivas} 
\begin{eqnarray}
T(t_2,t_0)=T(t_2,t_1)T(t_1,t_0).
\label{Eqn:divT}
\end{eqnarray}
In contrast to the non-Markovianity 
criteria in Refs. \cite{Rivas2010,Hou2011}, the violation the Eq.\ (\ref{Eqn:divT}) is manifested by the inequality 
\begin{eqnarray}
T(t_2,t_0)\neq T(t_2,t_1)T(t_1,t_0).
\label{Eqn:ineq}
\end{eqnarray}
By comparing the evolutions $\rho_S(t_2)=T(t_2,t_0)\rho_S(t_0)$ and $\rho'_S(t_2)=T(t_2,t_1)\rho_S(t_1)$ where the system states
at $t_1$ are both $\rho_S(t_1)=T(t_1,t_0)\rho_S(t_0)$, the fact $\rho_S(t_2)\neq \rho'_S(t_2)$ can be 
interpreted as the dependence of the system's future dynamics (from $t_1$ to $t_2$) on its history (from $t_0$ to $t_1$). 
On the other hand, in the process $\rho'_S(t_2)=T(t_2,t_1)T(t_1,t_0)\rho_S(t_0)$, the system's history (from $t_0$ to $t_1$) 
encoded in $\rho_{S\!E}(t_1)$ is erased by the process $\rho_{S\!E}(t_1)\rightarrow\rho_{S}(t_1)\otimes\rho_{E}(t_1)$ where $t_1=t_I$. 
Therefore, $\rho_S(t_2)\neq \rho'_S(t_2)$ signifies that the environment (as well as its correlations with
the system \cite{Rivas}) remembers the system's history which in turn influences the system's future.

The non-Markovnianity of is defined as \cite{Hou2015}
\begin{eqnarray}
N_M &=& \max_{t_1,t_2}D[\rho_{T(t_2,t_0)},\rho_{T(t_2,t_1)T(t_1,t_0)}]. 
\label{Eqn:NM}
\end{eqnarray}
where $D(\rho_1,\rho_2)=\frac{1}{2}\|\rho_1-\rho_2\|$ is the trace distance
with $\|A\|=\Tr(\sqrt{A^\dag A})$ the trace norm of an operator $A$, and $\rho_{\Lambda}$
represents the Choi-Jami\'{o}{\l}kowski matrices \cite{Choi,Jamiolkowski} of a dynamical map $\Lambda$.
It is given by $\rho_\Lambda=(\mathbb{I}\otimes\Lambda)(\ket{\psi_{A\!S}}\bra{\psi_{A\!S}})$ where $\mathbb{I}$ is 
the identity map and $\ket{\psi_{A\!S}}=\frac{1}{\sqrt{d}}\sum_{i=1}^{d}\ket{i}_A\ket{i}_S$ is a 
maximally entangled state between an ancillary system $A$ and the system $S$ (both $d$-dimensional).
Using dynamical maps $T(t_b,t_a)$ in its definition,  the measure does not depend 
on the system's initial state. To investigate the influence of particular initial states 
on the memory effects, we recently use  
\begin{eqnarray}
N_M[\rho_S(t_0)]=\max_{t_1,t_2} D[\rho_S(t_2),\rho'_S(t_2)]
\label{Eqn:NMI}
\end{eqnarray}
as the strength of memory effects conditioned on the initial state $\rho_S(t_0)$ \cite{Hou2024}.
Note that $N_M[\rho_S(t_0)]>0$ is a sufficient condition for the inequality  Eq.(\ref{Eqn:ineq}) or $N_M>0$. 
In the following, the above two measures are both considered to deal with the control of 
both dynamical maps and quantum states by dynamical decoupling.

\subsection{Memory effects with specific time instants}

Intuitively, there exists some special time instants in a dynamical decoupling process that are closely related
to the environmental memory effects. For example, in one cycle of parity kicks described by Eq.\ (\ref{Eqn:PKCycle}), 
the information about the system's history in $[0,\tau]$ [stored in $\rho_{S\!E}(\tau)$] is exploited to time-reverse 
the dynamics after the first kick and ultimately eliminate the influence of environment at $2\tau$ (after the second kick). 
This motivates us to focus on the memory effects associated with specific time instants (e.g., $t_1=\tau$ and $t_2=2\tau$ in a 
parity kick cycle), rather than maximize the memory effects over all the possible $t_1$ and $t_2$ in a time interval like
$[0,\infty]$ as in Ref.\ \cite{Hou2015} or a finite one $[0,t_{max}]$ as in Ref.\ \cite{Hou2024}. 
Such a treatment is expected to help us establish simple and quantitative connections between the dynamical decoupling 
characteristics and strength of memory effects. Moreover, the computational effort would be significantly reduced since 
the maximization over the time instants are not required.   

According to Eq.\ (\ref{Eqn:NM}), one can quantify the strength of memory effects (in terms of dynamical maps)
associated with two specific time instants $t_1$ and $t_2$ (let $t_0$ be fixed) simply by 
\begin{eqnarray}
N_M^{t_{1:2}}=D[\rho_{T(t_2,t_0)},\rho_{T(t_2,t_1)T(t_1,t_0)}],
\label{Eqn:NMt1t2}
\end{eqnarray}
regardless of the system initial state. Similarly, the strength of memory effects 
with specific $t_1$ and $t_2$ conditioned on the initial state $\rho_S(t_0)$ is 
given by  
\begin{eqnarray}
N_M^{t_{1:2}}[\rho_S(t_0)]=D[\rho_S(t_2),\rho'_S(t_2)].
\label{Eqn:NMIt1t2}
\end{eqnarray}
Obviously, the above two measures are special cases of Eq.\ (\ref{Eqn:NM}) and (\ref{Eqn:NMI})
with the same physical interpretations and satisfy $0\leqslant N_M^{t_{1:2}}\leqslant N_M\leqslant1$ and 
$0\leqslant N_M^{t_{1:2}}[\rho_S(t_0)]\leqslant N_M[\rho_S(t_0)]\leqslant1$. Any nonzero 
$N_M^{t_{1:2}}$ or $N_M^{t_{1:2}}[\rho_S(t_0)]$ is a sufficient condition for non-Markovianity. 
The strengths of memory effects defined by Eq.\ (\ref{Eqn:NMt1t2}) and (\ref{Eqn:NMIt1t2}) can 
be straightforwardly applied to Eq.\ (\ref{Eqn:PKCycle}) by letting $t_1=\tau$ (after the first kick) 
and $t_2=2\tau$ (after the second kick) to evaluate to what degree the system's final dynamical map 
$T(t_2,t_0)$ [or the final state $\rho_S(t_2)$] depends on its history before the first kick.

To evaluate the strength of memory effects in a dynamical decoupling process with more than 2 kicks, e.g.,
more than one cycle of Eq.\ (\ref{Eqn:PKCycle}), we extend Eq.\ (\ref{Eqn:NMt1t2}) and  (\ref{Eqn:NMIt1t2}) 
to their multi-time forms. In a Markovian quantum process, by repeatedly applying the divisibility 
Eq.\ (\ref{Eqn:divT}) to one of the terms on the right-hand side of Eq.\ (\ref{Eqn:divT}), there are
\begin{eqnarray}
\!\!\!\!\!\!\!\!\!\!T(t_n,t_0)=T(t_n,t_{n\!-\!1})\cdots T(t_2,t_1)T(t_1,t_0)
\label{Eqn:MTMdiv}
\end{eqnarray}
for $t_0\leqslant t_1\cdots\leqslant t_n$. Similarly, the violation of the multi-time divisibility condition 
is manifested by the inequality:
\begin{eqnarray}
T(t_n,t_0)\neq T(t_n,t_{n-1}) \cdots T(t_2,t_1) T(t_1,t_0).\quad
\label{Eqn:MTineq}
\end{eqnarray}
The multi-time inequality can also be interpreted as memory effects from the perspective of 
the environment as follows. Let the left-hand and the right-hand sides of Eq.\ (\ref{Eqn:MTineq}) 
act on a system initial state $\rho_S(t_0)$ so that $\rho_S(t_n)=T(t_n,t_0)\rho_S(t_0)$ and 
$\rho'_S(t_n)= T(t_n,t_{n-1}) \cdots T(t_2,t_1) T(t_1,t_0)\rho_S(t_0)$. 
The system-environment evolutions described by the latter equation can be understood as that by 
the former equation subjecting to $n-1$ initializations of the environment state at $t_{1,2,\cdots,n-1}$.
 The initialization at $t_j$ erases the system's history in the time interval $[t_{j-1}, t_j]$ 
 by the process $\rho_{S\!E}(t_j)\rightarrow\rho_{S}(t_j)\otimes\rho_{E}(t_j)$ with $t_j=t_I$.
 Therefore, $\rho_S(t_n)\neq \rho'_S(t_n)$ signifies that 
the environment (as well as the system-environment correlations) remembers the system's histories 
which in turn influences the future system at $t_n$. Then, the strength of memory effects (in terms
 of dynamical maps) with $n$ specific time instants $t_{1,2,\cdots, n}$ ($t_0$ is fixed) is given by 
\begin{eqnarray}
\!\!\!\!\!\!\!\!\!\!\!\!N_M^{t_{1:n}}
=D[\rho_{T(\!t_n,t_0\!)},\rho_{T(t_n,t_{n\!-\!1}) \cdots T(t_2,t_1) T(t_1,t_0)}].
\label{Eqn:NMt1tn}
\end{eqnarray}
Similarly, the strength of memory effects with specific $t_{1,2,\cdots, n}$   
conditioned on a particular system initial state $\rho_S(t_0)$ is given by  
\begin{eqnarray}
N_M^{t_{1:n}}[\rho_S(t_0)]=D[\rho_S(t_n),\rho'_S(t_n)]
\label{Eqn:NMIt1tn}
\end{eqnarray}
where $\rho_S(t_n)$ and $\rho'_S(t_n)$ are obtained from Eq.\ (\ref{Eqn:MTineq}) as mentioned before.
Then the strengths of memory effects with many specific time instants can be applied to a 
general dynamical decoupling processes described by one or more cycles of Eq.\ (\ref{Eqn:GeneralDD}) 
by letting $t_0=0$, and $t_j=j\tau$ (the time instant after each kick).  Within the definitions 
Eq.\ (\ref{Eqn:NMt1tn}) and (\ref{Eqn:NMIt1tn}), it is easy to see that $0\leqslant N_M^{t_{1:n}}\leqslant1$ and 
$0\leqslant N_M^{t_{1:n}}[\rho_S(t_0)]\leqslant1$ are satisfied. However, the values of $N_M^{t_{1:2}}$ and 
$N_M^{t_{1:n}}$ may not be compared directly since the number of environment initializations 
are different in the two measures, so do $N_M^{t_{1:2}}[\rho_S(t_0)]$ and $N_M^{t_{1:n}}[\rho_S(t_0)]$. 

\section{Memory effects and dynamical decoupling}
\label{Sec:MemoDD} 

It is known that a non-Markovian environment is necessary for the success of a dynamical decoupling 
control of an open quantum system. Therefore, a system with weak memory effects may lead to failure of 
dynamical decoupling. On the other hand, in an ideal dynamical decoupling process where 
infinitely frequent ($t_c \rightarrow 0$) instantaneous kicks act on the system, the system may be perfectly
decoupled from the environment so that the controlled dynamics becomes effectively unitary and thus Markovian. 
Thus we may not find a monotonic relationship between the performance of dynamical decoupling and the 
strength of memory effects in the controlled dynamics. In the following, we consider the memory effects as well as  
performances in both uncontrolled and controlled process to investigate their quantitative relationship.

We assume a separable system-environment initial state of the form 
Eq.\ (\ref{Eqn:InitialCondition}) with a fixed environment initial state and 
an arbitrary system initial state.  Then any uncontrolled system evolutions 
starting from $t_a$ can be expressed  in terms of Eq.\ (\ref{Eqn:T}) as
 \begin{eqnarray}
  \!\!\!\!\!\!\!\! \bar{\rho}_S(t_b)&\!=\!&\bar{T}(t_b,t_a)\rho_S(t_a)  \nonumber\\
  &\!=\!&\Tr_E[e^{-iH_0(t_b-t_a)}\rho_S(t_a)\otimes\rho_E(t_a)e^{iH_0(t_b-t_a)}].
\label{Eqn:T0}
\end{eqnarray}
where $H_0$ and $\rho_E(t_a)$ are given by Eq.\ (\ref{Eqn:Hmt}) and (\ref{Eqn:EnvInitialState}). 
Similarly, the system evolution starting from $t_a$ under the controlled total Hamiltonian Eq.\ (\ref{Eqn:HmtControl}) is given by 
\begin{eqnarray}
  \!\!\!\!\!\!\!\! \tilde{\rho}_S(t_b)&=&\tilde{T}(t_b,t_a)\rho_S(t_a)  \nonumber\\
  &=&\Tr_E[U(t_b,t_a)\rho_S(t_a)\otimes\rho_E(t_a)U(t_b,t_a)^{\dag}]
\label{Eqn:T1}
\end{eqnarray}
where $U(t_b,t_a)=\mathcal{T}e^{-i\int_{t_a}^{t_b}H_c(t')dt'}$. For example, for a cycle of ideal parity kicks 
starting at $t_a=0$, there is $U(\tau,0)=U(2\tau,\tau)=K_Pe^{-iH_0\tau}$.  

Let $\mathcal{K}_j\rho=K_j\rho K_j^{\dag}$ where $\mathcal{K}_j$ represents the dynamical map describing 
the action of an ideal kick $K_j$ on a density matrix $\rho$. For example, for parity kicks, 
there is  $\mathcal{K}_j\rho=\mathcal{K}_P\rho=K_P\rho K_P^{\dag}$. It is seen that 
\begin{eqnarray}
\tilde{T}[j\tau,(j-1)\tau]=\mathcal{K}_j \bar{T}[j\tau,(j-1)\tau]
\label{Eqn:TcKT0}
\end{eqnarray}
for a dynamical decoupling process stating at $t=0$ where $j=1,2,\cdots,n$. 

Furthermore, we refer to 
\begin{eqnarray}
  E=D[\rho_{\tilde{T}(t_f,0)},\rho_{\bar{T}(t_f,0)}]
\label{Eqn:EffectMap}
\end{eqnarray}
as the effect of a dynamical decoupling control (from $0$ to $t_f=kt_c$)
on the dynamical map $T(t_f,0)$. It represents the difference between
 the controlled (practically attainable) dynamical map $\tilde{T}(t_f,0)$ and the 
 uncontrolled one $\bar{T}(t_f,0)$.  Correspondingly, the effect of a dynamical decoupling control 
on the final state is given by
\begin{eqnarray}
 E[\rho_S(0)] &=& D[\tilde{\rho}_S(t_f),\bar{\rho}_S(t_f)] 
\label{Eqn:EffectState}
\end{eqnarray}
where $\rho_S(0)$ is the initial state, $\tilde{\rho}_S(t_f)=\tilde{T}(t_f,0)\rho_S(0)$ is 
the controlled final state and $\bar{\rho}_S(t_f)=\bar{T}(t_f,0)\rho_S(0)$ is uncontrolled one. 
Within the notations introduced above, we will explore the quantitative connections between the 
characteristics of a dynamical decoupling task and the strengths of memory effects (in both 
controlled and uncontrolled evolutions).

\subsection{One cycle of parity kicks}
Let $\tilde{N}_M^{t_{1:2}}$  denote the strength of memory effects for one cycle of 
parity kicks where $t_1=\tau$ and $t_2=t_c=2\tau$ (after each kick) and $t_0=0$, i.e.,
\begin{eqnarray}
\tilde{N}_M^{t_{1:2}}  &=& D[\rho_{\tilde{T}(2\tau,0)},\rho_{\tilde{T}(2\tau,\tau)\tilde{T}(\tau,0)}]. 
\label{Eqn:NMt1t2PK}
\end{eqnarray}
In contrast, for the same process but without control, the strength of memory effects 
with the same time instants, denoted by $\bar{N}_M^{t_{1:2}}$, is given by 
\begin{eqnarray}
\bar{N}_M^{t_{1:2}}  &=& D[\rho_{\bar{T}(2\tau,0)},\rho_{\bar{T}(2\tau,\tau)\bar{T}(\tau,0)}]. 
\label{Eqn:NMt1t2PK0}
\end{eqnarray}
It is observed that if $\mathcal{K}_P$ commutes with $\bar{T}(t_b,t_a)$, i.e., 
\begin{eqnarray}
\mathcal{K}_P\bar{T}(t_b,t_a)=\bar{T}(t_b,t_a)\mathcal{K}_P, 
\label{Eqn:Commutation}
\end{eqnarray}
then the effect of two parity kicks on the dynamical map, i.e., 
$E=D[\rho_{\tilde{T}(2\tau,0)},\rho_{\bar{T}(2\tau,0)}]$, is bounded by 
$\tilde{N}_M^{t_{1:2}}$ and $\bar{N}_M^{t_{1:2}}$ as
 \begin{eqnarray}
 |\tilde{N}_M^{t_{1:2}}-\bar{N}_M^{t_{1:2}}|\leqslant  E \leqslant \tilde{N}_M^{t_{1:2}}+\bar{N}_M^{t_{1:2}},
 \label{Eqn:PKvsNM}
\end{eqnarray}
which is a main finding of our work. The condition Eq.\ (\ref{Eqn:PKvsNM}) can be derived with the help 
of Eq.\ (\ref{Eqn:TcKT0}) and the triangle inequality
 \begin{eqnarray}
D(\rho_1,\rho_2)\leqslant D(\rho_1,\rho_3) +D(\rho_2,\rho_3)
 \label{Eqn:triangle}
\end{eqnarray}
that the trace distance obeys. The derivation of is presented in Appendix\ \ref{Sec:AppendixA}. 
The above result shows that under certain conditions, the upper and lower bounds of the control effect 
(on the final dynamical map) is determined by the strength of memory effects in both the controlled 
and uncontrolled process. Meanwhile, the difference between the controlled and uncontrolled dynamical 
map is a manifestation of memory effects since $E=0$ if $\tilde{N}_M^{t_{1:2}}=\bar{N}_M^{t_{1:2}}=0$.

If we are interested in the quantum states rather than the dynamical maps in a cycle of
parity kicks, the bounds of the memory effect strength can be obtained following similar 
steps. As discussed above, the strength of memory effects for one cycle of parity kicks 
with the initial state $\rho_S(0)$ is given by 
\begin{eqnarray}
\tilde{N}_M^{t_{1:2}}[\rho_S(0)]&=&D[\tilde{\rho}_S(t_c),\tilde{\rho}'_S(t_c)]
\label{Eqn:NMIt1t2PK}
\end{eqnarray}
where $\tilde{\rho}_S(t_c)=\tilde{T}(2\tau,0)\rho_S(0)$ and $\tilde{\rho}'_S(t_c)=\tilde{T}(2\tau,\tau)\tilde{T}(\tau,0)\rho_S(0)$.
In the absence of control, the strength of memory effects with the same time instants is given by
\begin{eqnarray}
\bar{N}_M^{t_{1:2}}[\rho_S(0)]&=&D[\bar{\rho}_S(t_c),\bar{\rho}'_S(t_c)]
\label{Eqn:NMIt1t20}
\end{eqnarray}
where $\bar{\rho}_S(t_c)=\bar{T}(2\tau,0)\rho_S(0)$ and $\bar{\rho}'_S(t_c)=\bar{T}(2\tau,\tau)\bar{T}(\tau,0)\rho_S(0)$. 
Then the effect of control on the final states $E[\rho_S(0)] = D[\tilde{\rho}_S(t_c),\bar{\rho}_S(t_c)]$ satisfies 
\begin{eqnarray}
 &&|\tilde{N}_M^{t_{1:2}}[\rho_S(0)]-\bar{N}_M^{t_{1:2}}[\rho_S(0)]|\leqslant E[\rho_S(0)] \nonumber\\ 
 &&\leqslant \tilde{N}_M^{t_{1:2}}[\rho_S(0)]+\bar{N}_M^{t_{1:2}}[\rho_S(0)]
 \label{Eqn:PKvsNMstate}
\end{eqnarray}
if $\mathcal{K}_P$ commutes with $\bar{T}(t_b,t_a)$ (see Appendix\ \ref{Sec:AppendixA} for its derivation). 
Similarly, the change of the final states (brought by control) is upper and lower bounded by the memory 
effects in the controlled and uncontrolled process. Meanwhile, $E[\rho_S(0)]>0$ is a manifestation 
of memory effects.

\subsection{General case}
Our theory can be extended to a dynamical decoupling process with $n$ kicks ($n\geqslant2$), 
specifically, one or more cycles of Eq.\ (\ref{Eqn:GeneralDD}) where the kicks are designed by 
Eq.\ (\ref{Eqn:Groupdesign}). Consider a dynamical decoupling process (from 0 to $t_f$) with $k$ 
cycles while each cycle contains $n_g$ kicks. When $k>1$, the kicks satisfies  $K_{j+n_g}=K_j$.  
The process can be described by $\tilde{T}(n\tau,0)$ where $n=kn_g$. Let $t_0=0$ and $t_j=j\tau$ 
(after each kick), then the strength of memory effects in the dynamical decoupling process satisfies  
\begin{eqnarray}
\!\!\!\!\!\!\!\!\!\!\!\!\tilde{N}_M^{t_{1:n}}
\!&=&\!D\{\rho_{\tilde{T}(n\tau,0)},\rho_{\tilde{T}[n\tau,(n\!-\!1)\tau] \cdots \tilde{T}(2\tau,\tau) \tilde{T}(\tau,0)}\} 
\label{Eqn:NMt1tnDD}
\end{eqnarray}
following the definition Eq.\ (\ref{Eqn:NMt1tn}) 
Using the same time instants $t_{1,2,\cdots, n}$, the strength of memory effects without 
control is given by 
\begin{eqnarray}
\!\!\!\!\!\!\!\!\!\!\!\!\bar{N}_M^{t_{1:n}}
\!&=&\!D\{\rho_{\bar{T}(n\tau,0)},\rho_{\bar{T}[n\tau,(n\!-\!1)\tau] \cdots \bar{T}(2\tau,\tau) \bar{T}(\tau,0)}\}.
\label{Eqn:NMt1tn0}
\end{eqnarray}
With the condition Eq.\ (\ref{Eqn:condition}) and the inequality Eq.\ (\ref{Eqn:triangle}),
it is found that if each $\mathcal{K}_j$ commutes with $\bar{T}(t_b,t_a)$, then 
 the effect of control on the dynamical map satisfies 
\begin{eqnarray}
 |\tilde{N}_M^{t_{1:n}}-\bar{N}_M^{t_{1:n}}|\leqslant  E \leqslant \tilde{N}_M^{t_{1:n}}+\bar{N}_M^{t_{1:n}}
 \label{Eqn:DDvsNM}
\end{eqnarray}
where $E=D[\rho_{\tilde{T}(n\tau,0)},\rho_{\bar{T}(n\tau,0)}]$. The derivation of 
Eq.\ (\ref{Eqn:DDvsNM}) is presented in Appendix\ \ref{Sec:AppendixB}. The result reduces to 
Eq.\ (\ref{Eqn:PKvsNM}) when $n=2$ ($k=1$ and $n_g=2$) and can be applied 
to more than one cycle of parity kicks ($k>1$ and $n_g=2$). If $n>2$, the interpretation
of $\tilde{N}_M^{t_{1:n}}$ ($\bar{N}_M^{t_{1:n}}$) as memory effects is slightly different with 
that of $\tilde{N}_M^{t_{1:2}}$ ($\bar{N}_M^{t_{1:2}}$) as discussed in Sec.\ \ref{Sec:Measure} B.

Similarly, the strength of memory effects for a general dynamical decoupling process 
with the initial state $\rho_S(0)$ is given by  
\begin{eqnarray}
\tilde{N}_M^{t_{1:n}}[\rho_S(0)]&=&D[\tilde{\rho}_S(kt_c),\tilde{\rho}'_S(kt_c)]
\label{Eqn:NMIt1tnDD}
\end{eqnarray}
where $\tilde{\rho}_S(kt_c)=\tilde{T}(n\tau,0)\rho_S(0)$ and $\tilde{\rho}'_S(kt_c)=\tilde{T}[n\tau,(n-1)\tau] 
\cdots \tilde{T}(2\tau,\tau)\tilde{T}(\tau,0)\rho_S(0)$. In the uncontrolled process, the strength of memory effects is
\begin{eqnarray}
\bar{N}_M^{t_{1:n}}[\rho_S(0)]&=&D[\bar{\rho}_S(kt_c),\bar{\rho}'_S(kt_c)]
\label{Eqn:NMIt1tn0}
\end{eqnarray}
where $\bar{\rho}_S(kt_c)=\bar{T}(n\tau,0)\rho_S(0)$ and 
$\bar{\rho}'_S(t_c)=\bar{T}[n\tau,(n-1)\tau]\cdots \bar{T}(2\tau,\tau)\bar{T}(\tau,0)\rho_S(0)$. 
At last, the effect of a general dynamical decoupling control on the final state, i.e.,
$E[\rho_S(0)] = D[\tilde{\rho}_S(kt_c),\bar{\rho}_S(kt_c)]$, satisfies 
\begin{eqnarray}
 &&|\tilde{N}_M^{t_{1:n}}[\rho_S(0)]-\bar{N}_M^{t_{1:n}}[\rho_S(0)]|\leqslant E[\rho_S(0)] \nonumber\\ 
 &&\leqslant \tilde{N}_M^{t_{1:n}}[\rho_S(0)]+\bar{N}_M^{t_{1:n}}[\rho_S(0)]
 \label{Eqn:DDvsNMstate}
\end{eqnarray}
if each $\mathcal{K}_j$ commutes with $\bar{T}(t_b,t_a)$ (see Appendix\ \ref{Sec:AppendixB} for its derivation). 

The general condition (\ref{Eqn:DDvsNM}) ($n\geqslant2$) can be experimentally verified by quantum process 
tomography \cite{Nielsen} of the corresponding dynamical maps. If $\mathcal{K}_j$ commutes with 
$\bar{T}(t_b,t_a)$, there is $\tilde{T}[n\tau,(n\!-\!1)\tau] \cdots \tilde{T}(2\tau,\tau) \tilde{T}(\tau,0)=
\bar{T}[n\tau,(n\!-\!1)\tau] \cdots \bar{T}(2\tau,\tau) \bar{T}(\tau,0)$ (see Appendix\ \ref{Sec:AppendixB}). 
Then the characterizations of $\bar{T}(n\tau,0)$, $\bar{T}[n\tau,(n\!-\!1)\tau] \cdots \bar{T}(2\tau,\tau) 
\bar{T}(\tau,0)$ and $\tilde{T}(n\tau,0)$ are adequate to calculate the Choi-Jami\'{o}{\l}kowski matrices in 
Eq.\ (\ref{Eqn:DDvsNM}). Typically, this requires the tomography of $n+2$ dynamical maps. 
Alternatively, using $\ket{\psi_{A\!S}}$ (defined in Sec.\ \ref{Sec:MeasureA}) as the initial state,  
$\rho_{\Lambda}$ can be obtained by quantum state tomography of the final state of the $A\!-\!S$ composite system 
after $\mathbb{I}\otimes\Lambda$ \cite{Wölk}. Similarly, the condition (\ref{Eqn:DDvsNMstate}) can be experimentally 
verified by quantum state tomography of the final states, $\bar{\rho}_S(kt_c)$, $\bar{\rho}'_S(kt_c)$ and 
$\tilde{\rho}_S(kt_c)$ while a particular initial state $\rho_S(0)$ is used. These final states can also be calculated
by their corresponding dynamical maps obtained from quantum process tomography. In the above cases, if $\rho_E(t_a)$ is 
stationary under $H_E$ (assuming $H_0$ is time-dependent), there is $\bar{T}(t_b,t_a)=\bar{T}(t_b-t_a,0)$ \cite{Chruscinski}. 
Then the characterization of $\bar{T}[n\tau,(n\!-\!1)\tau] \cdots \bar{T}(2\tau,\tau) \bar{T}(\tau,0)$ only
requires the tomography of $\bar{T}(\tau,0)$.  

The existence of the commutation relation between $K_j$ and  $\bar{T}(t_b,t_a)$ relies on the 
kick $K_j$ as well as the dynamical map $\bar{T}(t_b,t_a)$ which is a function of the system-environment Hamiltonian
and the environment initial state.  In general, the commutation relation can be verified with the analytical and numerical 
solution of the open dynamics, or by experiments. In Sec.\ \ref{Sec:CommCondition}, we propose several 
properties for $H_E$, $H_{S\!E}$ and $\rho_E(t_I)$ that guarantee the commutation relation between a parity kick 
$K_P$ and $\bar{T}(t_b,t_a)$, which is proved with the parities of the total excitation numbers in the ket and 
bra vectors of the environment initial states. However, general conditions that guarantee 
$K_j\bar{T}(t_b,t_a)=\bar{T}(t_b,t_a)K_j$ for an arbitrary dynamical decoupling process are still open.

\subsection{Effect and performance of control}
Let us further discuss the meaning of $E$ and $E[\rho_S(0)]$ mentioned above. Consider that the target of a 
dynamical decoupling control is to realize a particular type of dynamical maps or quantum states at $t_f=kt_c$.  
In this work, our target dynamical map $T_{\mathrm{tar}}(t_f,0)$  and target state $T_{\mathrm{tar}}(t_f,0)$
at $t_f$ are defined by 
\begin{eqnarray}
\rho_{\mathrm{tar}}(t_f)&=&T_{\mathrm{tar}}(t_f,0)\rho_S(0)\nonumber\\
&=&e^{-iH_St_f}\rho_S(0)e^{iH_St_f}
\label{Eqn:Target}
\end{eqnarray} 
which are the unitary operation governed by $H_S$ and the corresponding final state conditioned on the 
initial state $\rho_S(0)$. In ideal cases, e.g., continuous ideal kicks, or a proper total Hamiltonian 
$H_0$ with ideal kicks that leads to perfect decoupling for any $t_c$, there are $\tilde{T}(t_f,0)=T_{\mathrm{tar}}(t_f,0)$ and
 $\tilde{\rho}_S(t_f)=\rho_{\mathrm{tar}}(t_f)$. Then the effect of control can be interpreted as 
the decrease of the distance between $T(t_f,0)$  and $T_{\mathrm{tar}}(t_f,0)$  or that between 
$\rho_S(t_f)$ and $\rho_{\mathrm{tar}}(t_f)$ brought by the control.  In other words, the effect of 
control represents the increase of the performance defined by 
\begin{eqnarray}
&&P=1-D[\rho_{T(t_f,0)},\rho_{T_{\mathrm{tar}}(t_f,0)}]
\label{Eqn:PerfMap}\\
\mathrm{or}\ \ &&P[\rho_S(0)]=1-D[\rho_S(t_f),\rho_{\mathrm{tar}}(t_f)]
\label{Eqn:PerfState}
\end{eqnarray}
brought by the control where $\rho_{T(t_f,0)}$ [$\rho_S(t_f)$] represents the final dynamical 
map (final state) without or with control. For example, for perfect decoupling, $\Delta P=\tilde{P}-\bar{P}=E$
where $\tilde{P}=1-D[\rho_{\tilde{T}(t_f,0)},\rho_{T_{\mathrm{tar}}(t_f,0)}]=1$ is the performance 
with control, and $\bar{P}=1-D[\rho_{\bar{T}(t_f,0)},\rho_{T_{\mathrm{tar}}(t_f,0)}]$ is that 
without control.  The interpretation approximately holds 
if $\tilde{T}(t_f,0) \approx T_{\mathrm{tar}}(t_f,0)$ and 
$D[\rho_{\tilde{T}(t_f,0)},\rho_{\bar{T}(t_f,0)}]\gg D[\rho_{\tilde{T}(t_f,0)},\rho_{T_{\mathrm{tar}}(t_f,0)}]$. 
In terms of quantum states, these conditions are  $\tilde{\rho}_S(t_f) \approx \rho_{\mathrm{tar}}(t_f)$ and
 $D[\tilde{\rho}_S(t_f), \bar{\rho}_S(t_f)]\gg D[\tilde{\rho}_S(t_f), \rho_{\mathrm{tar}}(t_f)]$. 
 In the case of ineffective dynamical decoupling, i.e., $\tilde{T}(t_f,0) \neq T_{\mathrm{tar}}(t_f,0)$ 
 or $\tilde{\rho}_S(t_f) \neq \rho_{\mathrm{tar}}(t_f)$  the above interpretation may still 
 hold if the control makes the uncontrolled final dynamical map (state) move toward the target one.  

Besides its relationship with the control performance, 
 any nonzero $E$ or $E[\rho_S(0)]$ is a manifestation of non-Markovianity (under certain conditions). 
If Eq.\ (\ref{Eqn:DDvsNM}) or (\ref{Eqn:DDvsNMstate}) can be reduced to some simpler forms, 
the effect $E$ or $E[\rho_S(0)]$ may directly represent the strength of 
memory effects in the controlled or uncontrolled dynamics.

\section{Example}
\label{Sec:Example} 
In this section, we propose a class of system-environment Hamiltonians with certain environment 
initial states that support the commutation relation between a parity kick and the dynamical map without control. 
After that, we illustrate our theory with a specific example in which the qubit in a dissipative quantum Rabi model is 
controlled by parity kicks in different regimes. 

\subsection{Commutation relation between $\mathcal{K}_P$ and $\bar{T}(t_b,t_a)$ for a class of models}
\label{Sec:CommCondition} 
To investigate the role of memory effects in parity kicks with our theory, 
$\mathcal{K}_P\bar{T}(t_b,t_a)=\bar{T}(t_b,t_a)\mathcal{K}_P$ is required. The existence of this relation  
depends on the properties of the parity kick given by Eq.\ (\ref{Eqn:ParityKickConditon1}) and (\ref{Eqn:ParityKickConditon2}), 
and that of the dynamical map $\bar{T}(t_b,t_a)$ which is a function of $H_0 = H_S+H_E+H_{S\!E}$ and 
the environment initial state $\rho_E(t_a)=\rho_E(t_I)$ defined in Eq.\ (\ref{Eqn:EnvInitialState}).
Assume that the environment consists of $m$ subsystems ($m\geqslant1$) whose initial state has the 
form $\rho_E(t_a)=\sum \rho_{\mathbf{n}\mathbf{n'}}\ket{\mathbf{n}}\bra{\mathbf{n'}}$
where $\ket{\mathbf{n}}=\ket{n_1,n_2\cdots n_m}$ ($\ket{\mathbf{n}'}=\ket{n_1',n_2'\cdots n_m'}$) 
and $n_{j=1,2,\cdots m}\geqslant0$ is the excitation number of the $j$th subsystem.  When
$\ket{\mathbf{n}}=\ket{\mathbf{n}'}$, $\rho_{\mathbf{n}\mathbf{n}'}$ represents the diagonal
element, otherwise, when $\langle{\mathbf{n}}\ket{\mathbf{n}'}=0$, $\rho_{\mathbf{n}\mathbf{n'}}$ represents 
the off-diagonal element. We consider a class of models with the following properties.

(1)  At $t_a=0$, for every nonzero matrix element $p_{\mathbf{n}\mathbf{n}’}$ in $\rho_E(t_a)$,
the total excitation numbers in $\ket{\mathbf{n}}$ and $\bra{\mathbf{n}'}$ always have the same 
parity, i.e., $n_{\mathrm{ket}}=\sum_{j=1}^{m}{n_j}$ and $n_{\mathrm{bra}}=\sum_{j=1}^{m}{n_j'}$ are both 
even or odd.  A typical example is a fully decohered state 
$\sum p_{\mathbf{n}\mathbf{n}}\ket{\mathbf{n}}\bra{\mathbf{n}}$ where $n_{\mathrm{ket}}=n_{\mathrm{bra}}$.

(2) The action of the environment Hamiltonian $H_{E}$ on $\ket{\mathbf{n}}$ turns $\ket{\mathbf{n}}$ into 
bases whose total excitation numbers have the same parity with $n_{\textrm{ket}}$ (or annihilate  $\ket{\mathbf{n}}$),
that is, $H_E$ preserves the parity of the total excitation number in $\ket{\mathbf{n}}$.  Consequently, 
the environment initial state after $t_a=0$, i.e., $\rho_E(t_a)=e^{-iH_Et_a}\rho_E(0)e^{iH_Et_a}$, also satisfies 
the first property. A typical example is a Hamiltonian $H_E$ that conserves the total excitation number.

(3) The action of the interaction Hamiltonian $H_{S\!E}$ on $\ket{\mathbf{n}}$ turns $\ket{\mathbf{n}}$ 
into bases whose total excitation numbers have the opposite parity with $n_{\textrm{ket}}$ (or annihilate $\ket{\mathbf{n}}$),
 that is, $H_{S\!E}$ changes the parity of the total excitation number in $\ket{\mathbf{n}}$. For example, 
 a Hamiltonian $H_{S\!E}$ that increases or decreases the total excitation number in $\ket{\mathbf{n}}$ by 1.

In the following, we show that $\mathcal{K}_P$ commutes with the dynamical map $\bar{T}(t_b,t_a)$ with the above properties.
We first divide the Hamiltonian $H_0$ into   $H_+=H_S+H_E$ and $H_-=H_{S\!E}$ such that 
\begin{eqnarray}
K_PH_+&=&H_+K_P
\label{Eqn:KPHP}\\
K_PH_-&=&-H_-K_P
\label{Eqn:KPHM}
\end{eqnarray} 
due to  Eq.\ (\ref{Eqn:ParityKickConditon1}) and (\ref{Eqn:ParityKickConditon2}). Meanwhile, 
$H_+$ ($H_-$) preserves (changes) the parity of the total excitation number in $\ket{\mathbf{n}}$.  Then the evolution 
of the system in the absence of control can be described by 
\begin{eqnarray}
\rho_{S}(t_b)&=&\bar{T}(t_b,t_a)\rho_S(t_a)\nonumber\\
&=&\sum_{\mathbf{n}''}\bra{\mathbf{n}''}U_0\rho_S(t_a)\otimes(\sum \rho_{\mathbf{n}\mathbf{n'}}\ket{\mathbf{n}}\bra{\mathbf{n'}})U_0^{\dag}\ket{\mathbf{n}''}\quad\quad
\label{Eqn:Proof1}
\end{eqnarray} 
where $U_0=I-i(H_++H_-)(t_b-t_a)-(H_++H_-)^2(t_b-t_a)^2/2+\cdots$ is the Taylor series of $e^{-iH_0(t_b-t_a)}$ and 
$\ket{\mathbf{n}''}=\ket{n_1'',n_2''\cdots n_m''}$. Moreover, $n_{\mathrm{ket}}$ and $n_{\mathrm{bra}}$ have the same parity.
Note that Eq.\ (\ref{Eqn:Proof1}) can be expressed as a summation of infinite terms, i.e., $\rho_S(t_b)=\sum_l \rho_S^{(l)}$. 
Any term $\rho_S^{(l)}$ satisfies 
\begin{eqnarray}
\rho_S^{(l)}\propto\bra{\mathbf{n}''}H_{\alpha}\!\cdots \!H_{\alpha}\rho_S(t_a)\otimes\ket{\mathbf{n}}\bra{\mathbf{n}'}H_{\alpha}\!\cdots\! H_{\alpha}\ket{\mathbf{n}''}\quad\quad
\label{Eqn:Proof2}
\end{eqnarray} 
where $\alpha$ represents $+$ or $-$ and the number of $H_{\alpha}$ in Eq.\ (\ref{Eqn:Proof2}) ranges from 0 to $\infty$. 
The coefficient in Eq.\ (\ref{Eqn:Proof2}) is ignored for simplicity now and later.

It can be verified that if the number of $H_-$ in Eq.\ (\ref{Eqn:Proof2}) is odd, then $\rho_S^{(l)}=0$.  For example, 
there are three $H_-$ in the term $H_-H_+\rho_S(t_a)\otimes\ket{\mathbf{n}}\bra{\mathbf{n}'}H_-H_-H_+$,  
See that the action of $H_-H_+$ on $\ket{\mathbf{n}}$ increases the number of excitations
 by $n_{\mathrm{odd}}=\pm1,\pm3,\pm5\cdots$, or annihilate it (bases after action denoted 
 by $\ket{\mathbf{n},+n_{\mathrm{odd}}}$).  In contrast, the action $H_-H_-H_+$ on $\bra{\mathbf{n}'}$ 
 increases the number of excitations by $n_{\mathrm{even}}=0,\pm2,\pm4\cdots$, or annihilated it 
 (bases after action denoted by $\bra{\mathbf{n}',+n_{\mathrm{even}}}$). The reason is that any number of $H_+$ 
and an even number of $H_-$ do not change the parity of the total excitation numbers in $\ket{\mathbf{n}}$ 
or $\bra{\mathbf{n}'}$. Then any term after the actions of $H_-H_+$ and $H_-H_-H_+$ satisfies
 $\bra{\mathbf{n}''}\hat{\rho}_S\otimes\ket{\mathbf{n},+n_{\mathrm{odd}}}\bra{\mathbf{n}',+n_{\mathrm{even}}}\mathbf{n}''\rangle=0$
 where $\hat{\rho}_S$ is transformed from $\rho_S(t_a)$.  Similarly, any term as Eq.\ (\ref{Eqn:Proof2}) 
 with an odd number of $H_-$ turns $\ket{\mathbf{n}}$ and $\bra{\mathbf{n}'}$ into bases with different 
 excitation numbers (due to different parities) or annihilate one or two of them. 
 Then, any term obtained by expanding Eq.\ (\ref{Eqn:Proof2}), if not annihilated, satisfies  
 \begin{eqnarray}
\bra{\mathbf{n}''}\hat{\rho}_S\otimes\ket{\mathbf{n},+n_{\mathrm{odd}}}\bra{\mathbf{n}',+n_{\mathrm{even}}}\mathbf{n}''\rangle&=&0,
\label{Eqn:Proof3A} \\
\mathrm{or}\ \bra{\mathbf{n}''}\hat{\rho}_S\otimes\ket{\mathbf{n},+n_{\mathrm{even}}}\bra{\mathbf{n}',+n_{\mathrm{odd}}}\mathbf{n}''\rangle&=&0.
\label{Eqn:Proof3}
\end{eqnarray} 
So $\rho_S^{(l)}=0$ is guaranteed. On the other hand, if the number of $H_-$ is even in Eq.\ (\ref{Eqn:Proof2}), there is   
\begin{eqnarray}
\!\!\!\!&&K_P H_{\alpha}\!\cdots \!H_{\alpha}\rho_S(t_a)\otimes\ket{\mathbf{n}}\bra{\mathbf{n}'}H_{\alpha}\!\cdots\! H_{\alpha}K_P\nonumber\\
&=&(-1)^{n_{\mathrm{even}}}H_{\alpha}\!\cdots \!H_{\alpha}K_P\rho_S(t_a)\otimes\ket{\mathbf{n}}\bra{\mathbf{n}'}K_PH_{\alpha}\!\cdots\! H_{\alpha}\nonumber\\
&=&H_{\alpha}\!\cdots \!H_{\alpha}K_P\rho_S(t_a)K_P\otimes\ket{\mathbf{n}}\bra{\mathbf{n}'}H_{\alpha}\!\cdots\! H_{\alpha}
\label{Eqn:Proof4}
\end{eqnarray} 
according to Eq.\ (\ref{Eqn:KPHP}) and (\ref{Eqn:KPHM}).  Now we know that every term in Eq.\ (\ref{Eqn:Proof1}) with an even
number of $H_-$ satisfies Eq.\ (\ref{Eqn:Proof4}) for any system initial state. By inserting Eq.\ (\ref{Eqn:Proof4}) back into 
Eq.\ (\ref{Eqn:Proof1}), $\mathcal{K}_P\bar{T}(t_b,t_a)=\bar{T}(t_b,t_a)\mathcal{K}_P$ is obtained.

The environment Hamiltonian  with property (2) can be constructed by $H_E=\sum_{n\geqslant1} H_{E_n}$ where each term $H_{E_n}$ 
consists of the product of a even number of ladder operators. For example, one or a collection of non-interacting (or interacting) 
harmonica oscillators. Similarly, the interaction Hamiltonian with property (3) can be constructed by 
$H_{S\!E}=\sum_{n\geqslant1} S_n\otimes E_n$ where each $E_n$ consists of the product of an odd number of ladder operators. 
For example,  the quantum Rabi interaction $\sigma^+a^{\dag}+\sigma^-a+\sigma^+a+\sigma^-a^{\dag}$  where the harmonic 
oscillator represents the environment.

\subsection{Model and control}
\label{Sec:ModelControl} 
In this subsection, we consider a quantum Rabi model where a qubit interacts with a 
dissipative harmonic oscillator in a Markovian reservoir at temperature $T$.  
The parity kicks are implemented on the qubit, i.e., the system (denoted by $S$), 
to protect it from the influence of the environment (the dissipative oscillator, denoted by $E$).
The Hamiltonian of the qubit and oscillator is given by 
\begin{eqnarray}
H_0 &=& H_S+H_E+H_{S\!E}\nonumber\\
    &=&\frac{1}{2}\omega_S\sigma_z+\omega_Ea^{\dag}a+ g(\sigma^+ + \sigma^-)(a^{\dag} + a).
\label{Eqn:HmtRabi}
\end{eqnarray}
where $\omega_S$ ($\omega_E$) is the transition frequency of the qubit  (oscillator), $\sigma_z$ is
 the Pauli $z$ matrix, $g$ is the coupling strength, $\sigma-$ ($\sigma^+$) is the lowering (raising) operator 
of the qubit  and $a$ ($a^{\dag}$) is the annihilation (creation) operator of the oscillator. 
The dynamics of the $\rho_{S\!E}$ without control is described by the standard Lindblad master equation
\begin{eqnarray}
\dot\rho_{S\!E}=\mathcal{L}(\rho_{S\!E})=-i[H_0,\rho_{S\!E}]+\mathcal{D}(\rho_{S\!E})
\label{Eqn:ME}
\end{eqnarray}
where $\mathcal{D}(\rho_{S\!E})=\gamma(\bar{n}+1)(a\rho_{S\!E}a^{\dag}-\frac{1}{2}a^{\dag}a\rho_{S\!E}-\frac{1}{2}\rho_{S\!E}a^{\dag}a)
+\gamma\bar{n}(a^{\dag}\rho_{S\!E}a-\frac{1}{2}aa^{\dag}\rho_{S\!E}-\frac{1}{2}\rho_{S\!E}aa^{\dag})$.
Here $\gamma$ is the dissipation rate of the oscillator and $\bar{n}=1/(e^{\beta\omega_E}-1)$ is the 
mean excitation number of the oscillator at temperature $T$ where $\beta=\frac{1}{k_BT}$.  
As discussed before, we consider evolutions starting from the condition $\rho_{S\!E}(t_I)=\rho_S(t_I)\otimes\rho_E(t_I)$
 where  $\rho_S(t_I)$ and $t_I$ are arbitrary. Besides, we assume that $\rho_E(t_I)$ is governed by 
 $\rho_{E}(t_I)=e^{\mathcal{L}_Et}\rho_E(0)$ before an evolution where 
\begin{eqnarray}
\mathcal{L}_E(\rho_{E})=-i[H_E,\rho_{E}]+\mathcal{D}(\rho_{E}).
\label{Eqn:MEEn}
\end{eqnarray}
In this work, we assume that $\rho_E(0)$ is a thermal state at temperature $T$, then
 $\rho_E(t_I)$ is stationary with respect to Eq.(\ref{Eqn:MEEn}) and is given by  
\begin{eqnarray}
\rho_E(t_I)=\rho_E^{\mathrm{th}}=\sum_{n_0=0}^\infty\frac{\bar{n}}{(1+\bar{n})^{n_0+1}}\ket{n_0}\bra{n_0}.
\label{Eqn:ThermalState}
\end{eqnarray}

The parity kicks are chosen as $K_P=\sigma_{z}$, which is assumed to be ideal. It leads to an
instantaneous phase change in the qubit's quantum state. Starting from $t=0$, the controlled dynamics can
 be described by 
\begin{eqnarray}
\dot\rho_{S\!E}=\mathcal{L}_t(\rho_{S\!E})=-i[H_0+H_1(t),\rho_{S\!E}]+\mathcal{D}(\rho_{S\!E})
\label{Eqn:MEcontrol}
\end{eqnarray}
where $H_1(t)=\sum_{j=1}^{n}\delta(t-t_j)H_P$ that realizes $K_P=\sigma_{z}$ at $t_j=j\tau$. It is easy 
to seen that conditions Eq.\ (\ref{Eqn:ParityKickConditon1}) and (\ref{Eqn:ParityKickConditon2}) 
are satisfied for our scheme. Therefore, the qubit can be dynamically decoupled from its environment after one 
or more cycles of parity kicks if $t_c$ is short enough.  Starting from the condition 
$\rho_{S\!E}(0)=\rho_S(0)\otimes\rho_E^{\mathrm{th}}$ with arbitrary $\rho_S(0)$, 
the Choi-Jami\'{o}{\l}kowski matrix of the target final dynamical map $T_{\mathrm{tar}}(t_f,0)$ is given by   
\begin{eqnarray}
\!\!\!\!\!\!\!\!\rho_{T_{\mathrm{tar}}(t_f,0)}\!&=&\![\mathbb{I}\!\otimes \!T_{\mathrm{tar}}(t_f,0)] (\ket{\psi_{A\!S}}\bra{\psi_{A\!S}})\nonumber\\
\!&=&\! (\mathbb{I}\!\otimes\! e^{-iH_St_f})(\ket{\psi_{A\!S}}\bra{\psi_{A\!S}})(\mathbb{I}\otimes e^{iH_St_f})
\label{Eqn:ChoiTar}
\end{eqnarray} 
where  $t_f=kt_c$ and $\ket{\psi_{AS}}=\frac{1}{\sqrt{2}}(\ket{g}_A\ket{g}_S+\ket{e}_A\ket{e}_S)$
is the maximally entangled state of the ancillary system and our system. If we are interested in the
control of quantum states, the target final state is given by 
\begin{eqnarray}
\rho_{\mathrm{tar}}(t_f) =e^{-\frac{i}{2}\omega_S\sigma_zt_f}\rho_S(0)e^{\frac{i}{2}\omega_S\sigma_zt_f}.
\label{Eqn:ChoiTar}
\end{eqnarray} 
Similar control schemes have been theoretically and experimentally investigated 
with a damping JC model to study the coherence (decoherence) of the oscillator through the behaviors 
of the qubit \cite{Morigi,Meunier}.  These schemes amount to using one cycle of parity kicks while 
omitting the second kick since it does not change the interested  observable.

In the following, we discuss the commutation relation $\mathcal{K}_P\bar{T}(t_b,t_a)=\bar{T}(t_b,t_a)\mathcal{K}_P$
for this model. The evolution of the qubit in the absence of control could be understood as 
$\rho_{S}(t_b)=\bar{T}(t_b,t_a)\rho_S(t_a)=\Tr_{E,R}[e^{-iH_{S\!E\!R}(t_b-t_a)}\rho_S(t_a)\otimes\rho_E^{\mathrm{th}}\otimes
\rho_R^{\mathrm{th}}e^{iH_{S\!E\!R}(t_b-t_a)}]$  where  
\begin{eqnarray}
H_{S\!E\!R} &=& H_0 + H_R+H_{E\!R}\nonumber\\
    &=&H_0+ \sum_{j}\omega_jb_j^{\dag}b_j+ \sum_j(g_jab_j^{\dag}+g_j^*a^{\dag}b_j)
\label{Eqn:HmtRabiFull}
\end{eqnarray} 
represents the Hamiltonian of the qubit, the oscillator, a Markovian reservoir (denoted by R) 
and their interactions. The oscillator and reservoir constitute the full environment of the qubit.
The reservoir is in a thermal state at temperature $T$, which is given by  
\begin{eqnarray}
\rho_R^{\mathrm{th}}=\prod_{j=1}^{\infty}\sum_{n_j=0}^\infty\frac{\bar{n}(\omega_j)}{[1+\bar{n}(\omega_j)]^{n_j+1}}\ket{n_j}\bra{n_j}.
\label{Eqn:ThermalStateR}
\end{eqnarray}
Then the $E\!-\!R$ initial state can be concisely expressed as
\begin{eqnarray}
\rho_E^{\mathrm{th}}\otimes\rho_R^{\mathrm{th}}&=&\sum \rho_{\mathbf{n}\mathbf{n}}\ket{\mathbf{n}}\bra{\mathbf{n}}
\label{Eqn:ERStateInFock}
\end{eqnarray} 
where $\ket{\mathbf{n}}=\ket{n_0,n_1,n_2\cdots}$, $p_{\mathbf{n}\mathbf{n}}$ is the diagonal element 
in the basis $\ket{\mathbf{n}}$. In this case, for every nonzero matrix element in Eq.\ (\ref{Eqn:ERStateInFock}),
$n_{\mathrm{ket}}$ and $n_{\mathrm{bra}}$ always have the same parity since $n_{\mathrm{ket}}=n_{\mathrm{bra}}$.
Now we divide the $S\!-\!E\!-\!R$ Hamiltonian Eq.\ (\ref{Eqn:HmtRabiFull}) into $H_+=H_S+H_E+H_R+H_{E\!R}$ 
and $H_-=H_{S\!E}$ such that $\sigma_zH_+=H_+\sigma_z$ and $\sigma_zH_-=-H_-\sigma_z$. Note that $H_+$ 
conserves the total excitation numbers in $\ket{\mathbf{n}}$ ($\bra{\mathbf{n}}$) and thus preserves the 
parities of $n_{\mathrm{ket}}$ and $n_{\mathrm{bra}}$. In contrast, $H_-$ increases or decreases the total 
excitation number in $\ket{\mathbf{n}}$ ($\bra{\mathbf{n}}$) by 1 and thus changes the parities of 
$n_{\mathrm{ket}}$ and $n_{\mathrm{bra}}$. Therefore,  $\mathcal{K}_P\bar{T}(t_b,t_a)=\bar{T}(t_b,t_a)\mathcal{K}_P$
 holds for this model according to the analysis in Sec.\ \ref{Sec:CommCondition}. Besides, the commutation relation
can be straightforwardly verified by numerically simulating Eq.\ (\ref{Eqn:ME}) before and after the kick 
(with different qubit initial states), or directly calculating and comparing the Choi-Jami\'{o}{\l}kowski matrices 
 $\rho_{\mathcal{K}_P\bar{T}(t_b,t_a)}$ and $\rho_{\bar{T}(t_b,t_a)\mathcal{K}_P}$ by simulations. 
 
 Following similar analysis or through numerical simulations, it is able to verify that the commutation relation
 might also holds for more related models. For example, a dissipative (or undamped) JC model, our model where 
 the oscillator's initial state satisfies $\rho_E(t_I)=\sum_{n_0=0}^\infty p_{n_0}\ket{n_0}\bra{n_0}$ 
 (not limited to thermal states), a qubit coupled to multiple dissipative (or undamped) oscillators, 
 a qubit coupled to a set of interacting oscillators, etc. Moreover, as known from the pseudomode 
 theory \cite{Garraway1997A,Garraway1997B} and its recent progresses \cite{Trivedi,Pleasance,Tamascelli}, 
 the reduced dynamics of an open quantum system with a complex environment (infinite modes) is equivalent 
 to that with a simpler environment (finite modes each subject to a Lindblad evolution) under certain conditions. 
 Therefore, it is possible to apply our theory to more complex environments \cite{Albarelli} to study the role 
 of memory effects in parity kicks.  
 
\subsection{Numerical simulations}
The above analysis demonstrates that the commutation relation holds irrespective of the values of $\omega_S$, $\omega_E$, 
$g$, $\gamma$ and $\bar{n}$. Thus our theory applies to different regimes, for example, that with a strong or weak coupling 
between $S$ and $E$, that with a detuning or resonance condition between $S$ and $E$, those with different temperatures 
(characterized by $\bar{n}$) or those with different $g/\gamma$ (determining the non-Markovianity). In the following, we numerically 
simulate the open dynamics of the qubit and calculate the strengths of memory effects (with and without control), the effects of 
control, and the performances (with and without control) in several typical regimes. These results will illustrate how the
effects of parity kicks are bounded by the strengths of memory effects (with and without control), as well as whether the
effect of  control reflects the increase of performance.  

Specifically, the Choi-Jami\'{o}{\l}kowski matrix $\rho_{\bar{T}(t_b,t_a)}$ is given by 
\begin{eqnarray}
\!\!\!\!\!\!\!\!\!\!\!\!\rho_{\bar{T}(t_b,t_a)}\!&=&\![\mathbb{I}\otimes \bar{T}(t_b,t_a)](\ket{\psi_{A\!S}}\bra{\psi_{A\!S}})\nonumber\\
\!&=&\!\Tr_E[(\mathbb{I}\otimes e^{\mathcal{L}(t_b-t_a)})(\ket{\psi_{A\!S}}\bra{\psi_{A\!S}}\otimes\rho_E^{\mathrm{th}})]
\label{Eqn:ChoiCal}
\end{eqnarray} 
where $\mathcal{L}$ is defined in Eq.\ (\ref{Eqn:ME}) and $\ket{\psi_{AS}}=\frac{1}{\sqrt{2}}(\ket{g}_A\ket{g}_S+\ket{e}_A\ket{e}_S)$. 
To calculate $\rho_{\bar{T}(t_b,t_a)}$, we first extend Eq.\ (\ref{Eqn:ME}) to the $A\!-\!S\!-\!E$ space by 
replacing  $O_{S\!E}$ with $I_A\otimes O_{S\!E}$ where $I_A$ is the identity operator in the ancillary space and  $O_{S\!E}$ 
represents the operators in Eq.\ (\ref{Eqn:ME}). Then, using $\rho_{A\!S\!E}(t_a)=\ket{\psi_{A\!S}}\bra{\psi_{A\!S}}\!\otimes\!\rho_E^{\mathrm{th}}$ 
as the initial state, the extend master equation is solved by a fourth-order Runge-Kutta method from to get $\rho_{A\!S\!E}(t_b)$.
At last, the result is obtained by $\rho_{\bar{T}(t_b,t_a)}=\Tr_E[\rho_{A\!S\!E}(t_b)]$. In each simulation, the Fock space of the 
oscillator is truncated to $\ket{n_{\mathrm{max}}}$ where $n_{\mathrm{max}}$ is large enough to guarantee the convergence of all the results, 
The Choi-Jami\'{o}{\l}kowski matrix $\rho_{\tilde{T}(t_b,t_a)}$ is given by 
\begin{eqnarray}
\!\!\!\!\!\!\!\!\!\!\!\!\rho_{\tilde{T}(t_b,t_a)}\!&=&\![\mathbb{I}\otimes \tilde{T}(t_b,t_a)](\ket{\psi_{AS}}\bra{\psi_{AS}})\nonumber\\
\!&=&\!\Tr_E[(\mathbb{I}\!\otimes \!\mathcal{T} \!e^{\int_{t_a}^{t_b}\!\mathcal{L}_{t'} dt'})
(\ket{\psi_{AS}}\bra{\psi_{AS}}\!\otimes\!\rho_E^{\mathrm{th}})]
\label{Eqn:ChoiControlCal}
\end{eqnarray} 
where $\mathcal{L}_{t'}$ is defined in Eq.\ (\ref{Eqn:MEcontrol}). It is calculated with Eq.\ (\ref{Eqn:MEcontrol}) with 
similar steps as done for $\rho_{\bar{T}(t_b,t_a)}$ where the kicks are modeled as instantaneous. Moreover, to calculate the strength of the memory effects, the 
Choi-Jami\'{o}{\l}kowski matrix $\rho_{\tilde{T}[n\tau,(n\!-\!1)\tau] \cdots \tilde{T}(2\tau,\tau) \tilde{T}(\tau,0)}$ and 
$\rho_{\bar{T}[n\tau,(n\!-\!1)\tau] \cdots \bar{T}(2\tau,\tau) \bar{T}(\tau,0)}$ (being the same) are needed. 
They can be calculated by
\begin{eqnarray}
\!\!\!\rho_{\bar{T}(2\tau,\tau) \bar{T}(\tau,0)}\!&=&\![\mathbb{I}\otimes \bar{T}(2\tau,\tau)\bar{T}(\tau,0)](\ket{\psi_{AS}}\bra{\psi_{AS}})\nonumber\\
\!&=&\![\mathbb{I}\!\otimes\!\bar{T}(2\tau,\tau)]\rho_{\bar{T}(\tau,0)},
\label{Eqn:ChoiEraseCal1}\\
\!\!\!\rho_{\bar{T}(3\tau,2\tau) \bar{T}(2\tau,\tau)\bar{T}(\tau,0)}\!&=&\![\mathbb{I}\!\otimes\!\bar{T}(3\tau,2\tau)]\rho_{\bar{T}(2\tau,\tau) \bar{T}(\tau,0)},
\label{Eqn:ChoiEraseCal2}\\
&\cdots& \nonumber
\end{eqnarray} 
The above calculations can be further simplified by $\bar{T}[j\tau,(j\!-\!1)\tau)]=\bar{T}(\tau,0)$ \cite{Chruscinski} and the property
\begin{eqnarray}
\Lambda\rho=d(\Tr \otimes \mathbb{I})[(\rho^{\mathrm{T}}\otimes I)\rho_{\Lambda}]
\label{Eqn:propertyChoi}
\end{eqnarray} 
where $\rho^{\mathrm{T}}$ is the transposition of the density matrix $\rho$, $d$ is the dimension of the system and 
$\rho_\Lambda$ is the Choi-Jami\'{o}{\l}kowski matrix of a dynamical map $\Lambda$ defined in Sec.\ \ref{Sec:Measure}.
 Now the quantities $\tilde{N}_M^{t_{1:n}}$, $\bar{N}_M^{t_{1:n}}$, $E$ and $P$ for the dynamical maps can be calculated
straightforwardly from the definitions (\ref{Eqn:EffectMap}), (\ref{Eqn:NMt1tnDD}), (\ref{Eqn:NMt1tn0}) and (\ref{Eqn:PerfMap}). 
If we are interested in the final state controlled by the parity kicks when an initial state is prescribed, the strengths of
memory effects (\ref{Eqn:NMIt1tnDD}) and (\ref{Eqn:NMIt1tn0}), the effects of control (\ref{Eqn:NMIt1tn0}),  and the
performance (\ref{Eqn:PerfState}) can be calculated by numerically simulating the master equation Eq.\ (\ref{Eqn:ME}) 
and (\ref{Eqn:MEcontrol}) and tracing out the environment.  Alternatively, they can be calculated directly with the
help of Eq.\ (\ref{Eqn:propertyChoi}) through the Choi-Jami\'{o}{\l}kowski matrices 
Eq.\ (\ref{Eqn:ChoiCal})-(\ref{Eqn:ChoiEraseCal2}) and so on. For example, 
$\tilde{\rho}_S(t_c)=\tilde{T}(2\tau,0)\rho_S(0)=2(\Tr\otimes\mathbb{I})\{[\rho_S(0)^\mathrm{T}\otimes I] \rho_{\tilde{T}(2\tau,0)}\}$.

\subsection{Results for one cycle of parity kicks}

\begin{figure}
\includegraphics*[width=9.2cm]{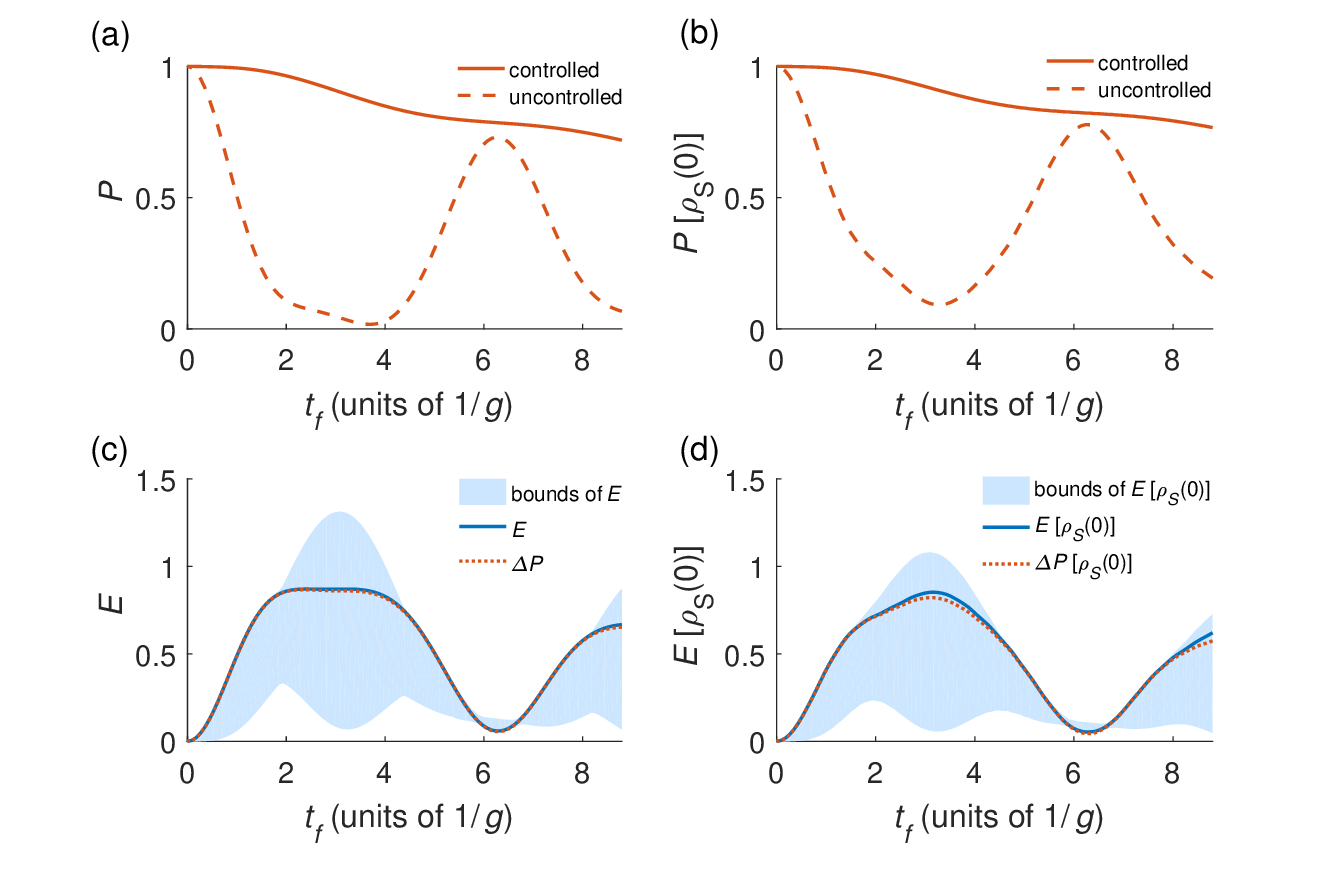}
\caption{Performances (with and without control), effects of control and their bounds as a function of the final time $t_f$ 
for 1 cycle of parity kicks ($t_f=t_c=2\tau$). The parameters are set as $\omega_E=\omega_S$, $g/\omega_S=0.01$, $\gamma/g=0.1$ and $\bar{n}=0.1$.
 (a) and (c) represent those for the final dynamical maps, while (b) and (d) for the final states. The system initial state
 in (b) and (c) is $1/\sqrt{2}(\ket{g}+\ket{e})$.  The difference of the two lines in (a) is plotted in (c) by the dotted 
 line for comparisons with $E$, similarly for (b) and (d).  }
\label{FIG:PicA}
\end{figure}

The role of memory effects in one cycle of parity kicks is examined as follows.  We first consider a regime where 
the qubit is weakly and resonantly coupled to the oscillator with a weak dissipation at a low temperature. 
The parameters are set as $g/\omega_S=0.01$, $\gamma/g=0.1$ and $\bar{n}=0.1$. The performance $P$ [defined by Eq.\ (\ref{Eqn:PerfMap})] 
with and without control are plotted in Fig.\ \ref{FIG:PicA}(a) as a function of the final time $t_f$ with $t_f=t_c=2\tau$. 
As shown in Fig.\ \ref{FIG:PicA}(a) by the solid line, when the cycle time is short, e.g., $t_c<1/g$, there is $P\approx1$, 
which means that the parity kicks can efficiently protect the qubit from the influence of the environment. Since the coupling 
strength $g$ is weak and $\omega_E=\omega_S$, the rotating wave approximation (RWA) works well here, i.e., 
the counter-rotating terms in Eq.\ (\ref{Eqn:HmtRabi}) do not influence the dynamics so much. Therefore, 
the decease of the controlled performance $P$ for larger $t_c$ is mainly due to the dissipation of the oscillator. 
In contrast, the performance in the absence of control (the dashed line) exhibits damped oscillations with respect
to $t_c$ and a fast decay when $0.5/g\lesssim t_c\lesssim1.5/g$. The effect of control $E$ as a function of $t_f$ is
plotted in Fig.\ \ref{FIG:PicA}(c) by the solid line, meanwhile, the bounds  
$E^{\mathrm{ub}}=\tilde{N}_M^{t_{1:2}}+\bar{N}_M^{t_{1:2}}$ and $E^{\mathrm{lb}}=|\tilde{N}_M^{t_{1:2}}-\bar{N}_M^{t_{1:2}}|$ 
are shown by the upper and lower edge of the shadow region. It is seen that the effect of control is upper and lower 
bounded by $E^{\mathrm{ub}}$ and $E^{\mathrm{lb}}$ as predicted by our theory. This property holds throughout our work, similar property also holds 
for the strength of memory effects in terms of quantum states. Specially, the relation between $E$ and the strengths of memory effects 
becomes quite simple in some cases. For example, when $t_c\lesssim1.5/g$, there is $E\approx\tilde{N}_M^{t_{1:2}}+\bar{N}_M^{t_{1:2}}$;
when $t_c\lesssim0.5/g$ where the dynamics and strengths of memory effects are almost quadratic with respect to $t_c$ \cite{Hou2024}, 
there is $|\tilde{N}_M^{t_{1:2}}-\bar{N}_M^{t_{1:2}}|\approx0$ such that $\tilde{N}_M^{t_{1:2}}\approx\bar{N}_M^{t_{1:2}}\approx2E$.
Besides, the difference of the controlled and uncontrolled performances ($\Delta P=\tilde{P}-\bar{P}$) is also plotted 
in Fig.\ \ref{FIG:PicA}(c) by the dotted line. In this regime, there is $\Delta P\approx E$ which means that the effect of control 
well reflects the increase of performance from the uncontrolled evolution to the controlled one, even when 
$\tilde{T}(t_f,0)\neq T_{\mathrm{tar}}(t_f,0)$. As done in Fig.\ \ref{FIG:PicA}(a) and (c), we also plot the performance
 $P[\rho_S(0)]$ with and without control, the effect of control $E[\rho_S(0)]$,  
$E^{\mathrm{ub}}[\rho_S(0)]=\tilde{N}_M^{t_{1:2}}[\rho_S(0)]+\bar{N}_M^{t_{1:2}}[\rho_S(0)]$ and 
$E^{\mathrm{lb}}[\rho_S(0)]=|\tilde{N}_M^{t_{1:2}}[\rho_S(0)]-\bar{N}_M^{t_{1:2}}[\rho_S(0)]|$ as a function of $t_f=t_c=2\tau$ in 
Fig.\ \ref{FIG:PicA}(b) and (d). The initial state is $\rho_S(0)=\ket{\phi}\bra{\phi}$ where $\ket{\phi}=1/\sqrt{2}(\ket{g}+\ket{e})$.
In general, the curves in Fig.\ \ref{FIG:PicA}(c) and (d) are different with those Fig.\ \ref{FIG:PicA}(a) and (b) since
they depend on the initial state $\rho_S(0)$. However, the two kinds of results are similar in many aspects since
the result in terms of quantum states is determined by that in terms of dynamical maps. For example, in Fig.\ \ref{FIG:PicA}(c) and (d), there is $\tilde{N}_M^{t_{1:2}}[\rho_S(0)]\approx\bar{N}_M^{t_{1:2}}[\rho_S(0)]\approx2E[\rho_S(0)]$ for $t_c\lesssim0.5/g$.

  \begin{figure}
\includegraphics*[width=9.2cm]{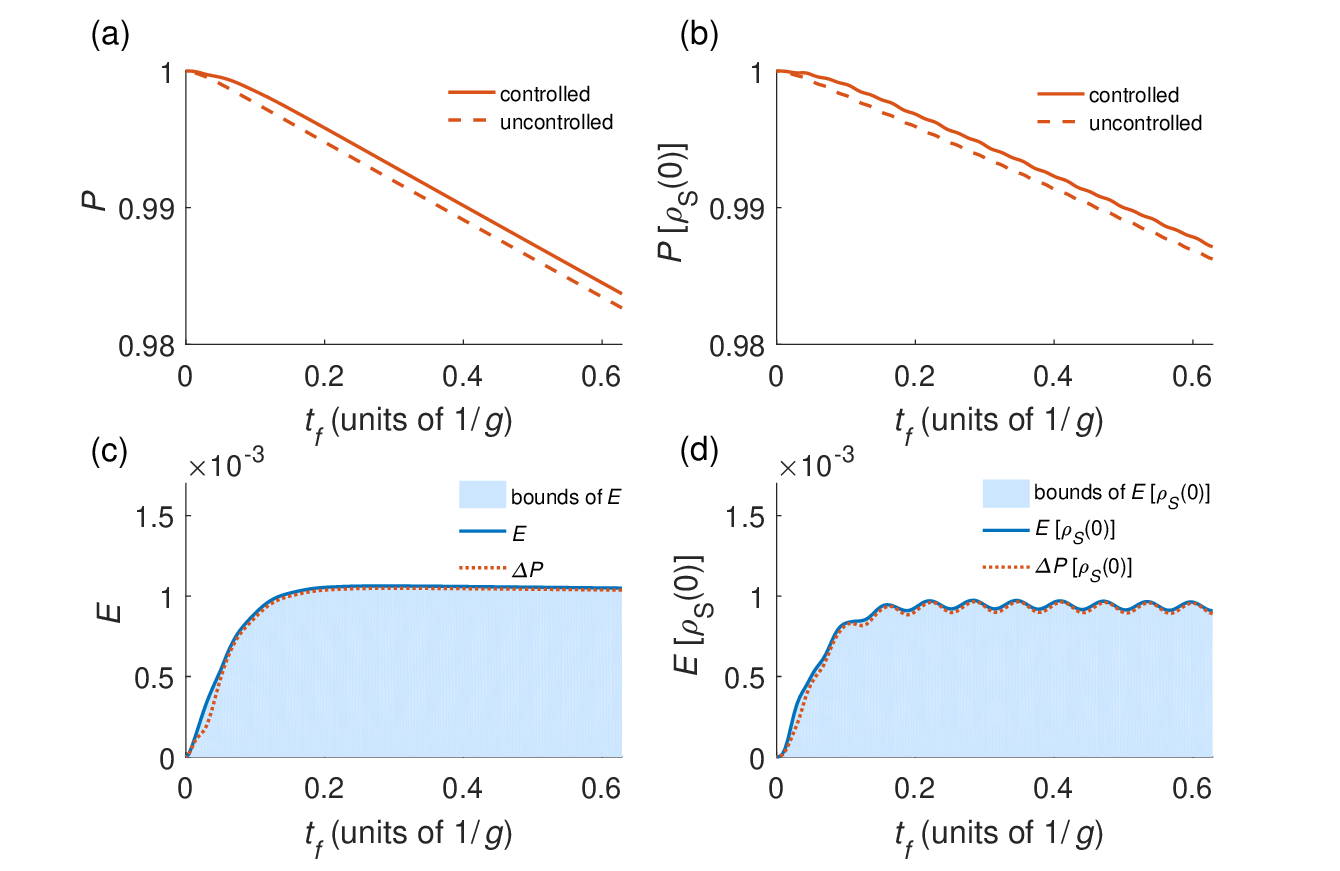}
\caption{Performances (with and without control), effects of control and their bounds as a function of the final time $t_f$ 
for 1 cycle of parity kicks. The parameters and settings are the same as those in Fig.\ \ref{FIG:PicA} except that $\gamma/g=100$.}
\label{FIG:PicB}
\end{figure}

 \begin{figure}
\includegraphics*[width=9.2cm]{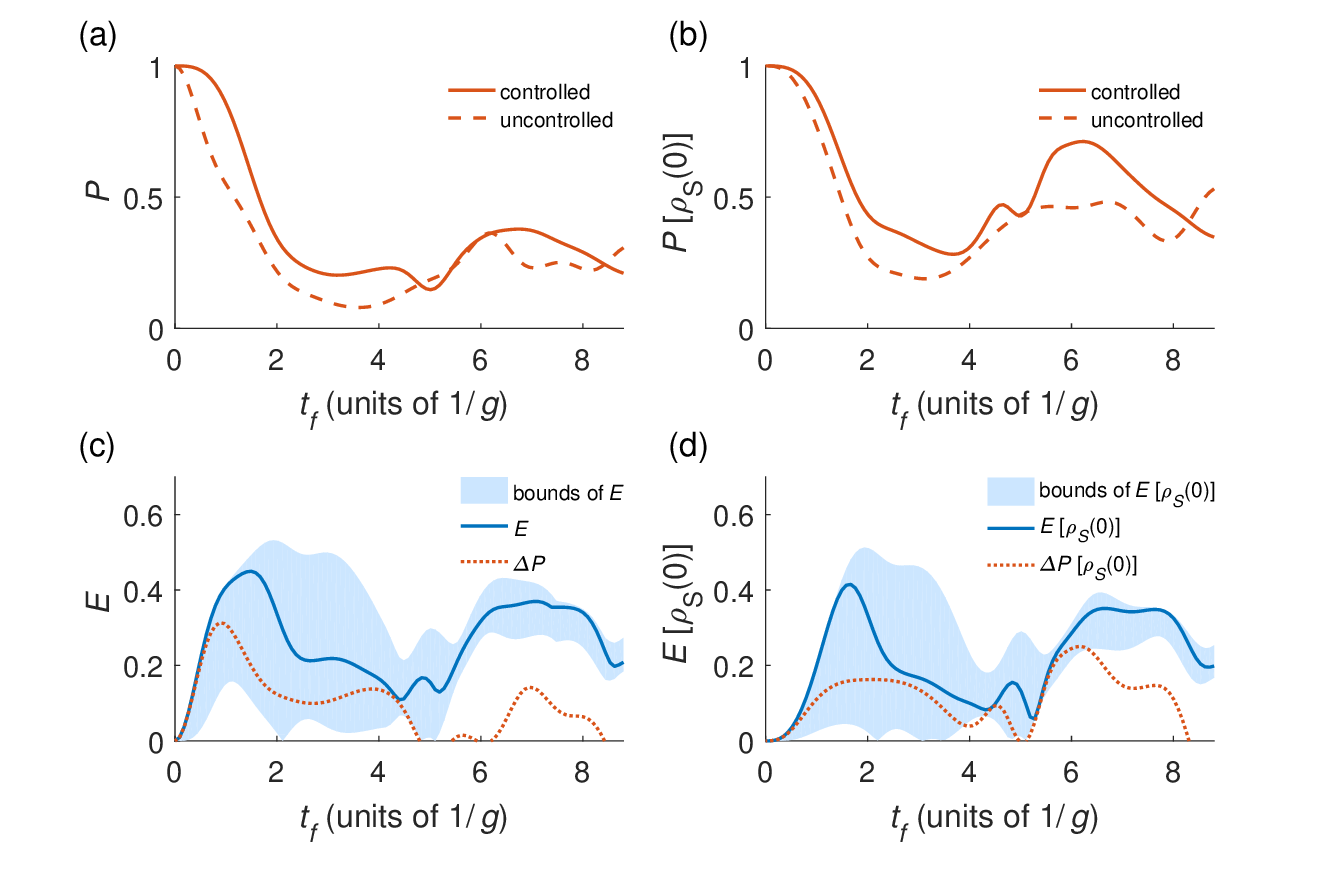}
\caption{Performances (with and without control), effects of control and their bounds as a function of the final time $t_f$ 
for 1 cycle of parity kicks. The parameters and settings are the same as those in Fig.\ \ref{FIG:PicA} except that $g/\omega_S=1$ and $\gamma/g=0.01$. }
\label{FIG:PicC}
\end{figure}

\begin{figure}
\includegraphics*[width=9.2cm]{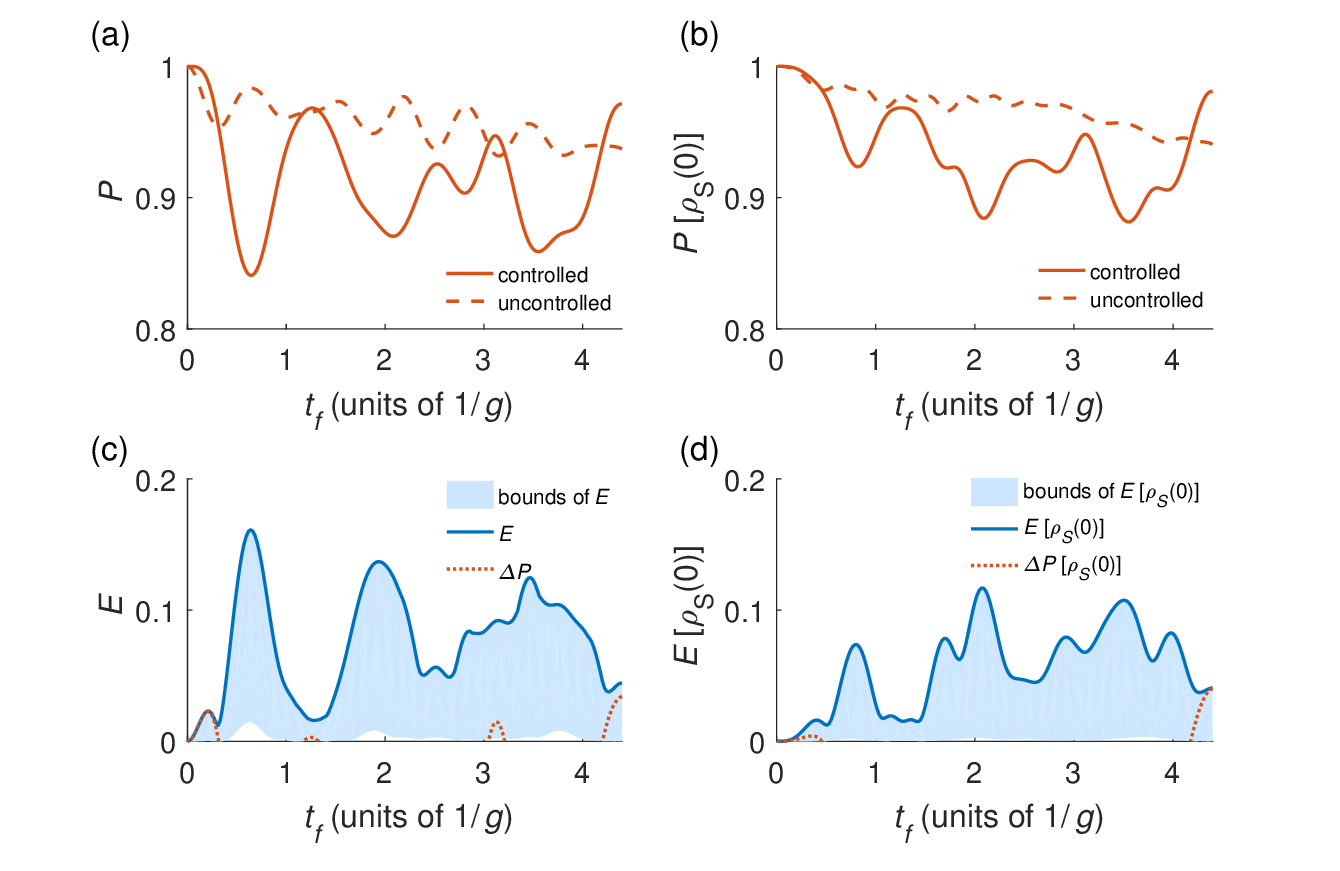}
\caption{Performances (with and without control), effects of control and their bounds as a function of the final time $t_f$ 
for 1 cycle of parity kicks. The parameters and settings are the same as those in Fig.\ \ref{FIG:PicC} except that $\omega_E=10\omega_S$. }
\label{FIG:PicD}
\end{figure}

We proceed to investigate  the role of memory effects in one cycle of parity kicks in several other regimes where the system 
initial states are the same as that in Fig.\ \ref{FIG:PicA}. The results for $\omega_E=\omega_S$, $g/\omega_S=0.01$, 
$\gamma/g=100$, $\bar{n}=0.1$ are shown in Fig.\ \ref{FIG:PicB}. The dynamics is highly Markovian compared with that in 
Fig.\ \ref{FIG:PicA} due to the strong dissipation of the oscillator as shown by the small values of the upper bounds of 
strengths of memory effects in Fig.\ \ref{FIG:PicB}(c) and (d).  Compared with the results in Fig.\ \ref{FIG:PicA}, 
the performances $P$ and $P[\rho_S(0)]$ without control decay monotonically with respect to $t_c$. Meanwhile, 
the parity kicks only result in a slight increase of the performances as shown in Fig.\ \ref{FIG:PicB}(a) and (b).
The reason is that the effects of control and further, the increase of performances, are limited by the weak memory effects. 
Besides, there are $\tilde{N}_M^{t_{1:2}}\approx\bar{N}_M^{t_{1:2}}\approx2E$, 
$\tilde{N}_M^{t_{1:2}}[\rho_S(0)]\approx\bar{N}_M^{t_{1:2}}[\rho_S(0)]\approx2E[\rho_S(0)]$, $\Delta P\approx E$ and 
$\Delta P[\rho_S(0)]\approx E[\rho_S(0)]$ for a wide range of $t_f$ as shown in Fig.\ \ref{FIG:PicB}(c) and (d). Further 
simulations shows that the fast and slight oscillations of $E[\rho_S(0)]$, $E^{\mathrm{ub}}[\rho_S(0)]$ and 
$\Delta P[\rho_S(0)]$ is caused by the counter-rotating terms in the qubit-oscillator interactions.

\begin{figure}
\includegraphics*[width=9.2cm]{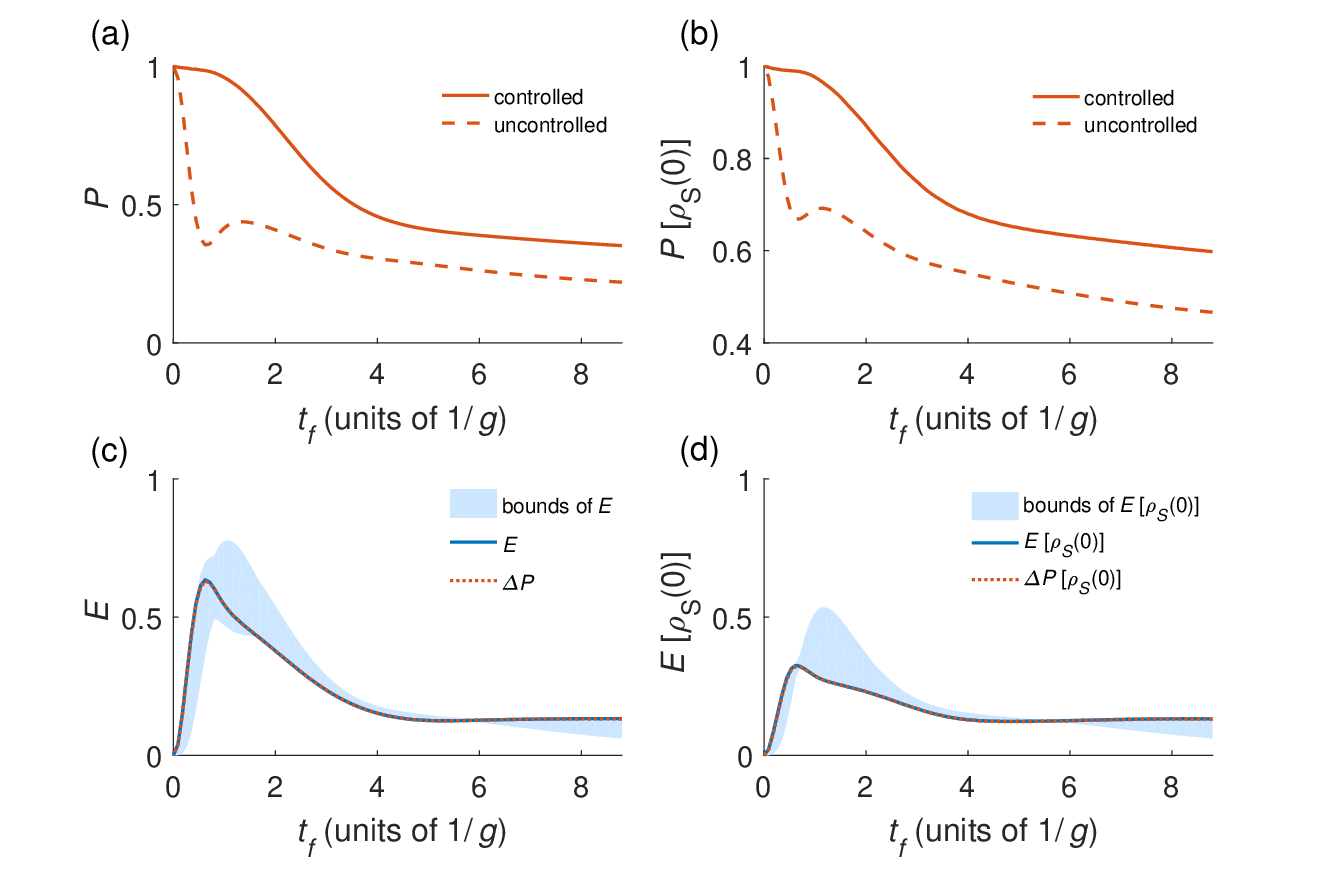}
\caption{Performances (with and without control), effects of control and their bounds as a function of the final time $t_f$ 
for 1 cycle of parity kicks. The parameters and settings are the same as those in Fig.\ \ref{FIG:PicA} except that $\bar{n}=5$.}
\label{FIG:PicAHeat}
\end{figure}

The results for $\omega_E=\omega_S$, $g/\omega_S=1$, $\gamma/g=0.01$, and $\bar{n}=0.1$ are shown in Fig.\ \ref{FIG:PicC}.
In contrast to the regime in Fig.\ \ref{FIG:PicA}, the qubit-oscillator coupling (characterized by $g/\omega_S$) is
much stronger and the rate $\gamma/g$ is lower.  So the dynamics is strongly influenced by the counter-rotating 
interactions rather than the dissipation in the time scale depicted in Fig.\ \ref{FIG:PicC}. It is observed that 
the performance $P$ and $P[\rho_S(0)]$ under control decay faster with respect to $t_f$ when $0.5/g\lesssim t_c\lesssim2/g$ compared 
with those in Fig.\ \ref{FIG:PicA}. Meanwhile, the dependence of the performances, the effects of control, and their
bounds on $t_c$ becomes complex compared with that in Fig.\ \ref{FIG:PicA}. For some values of $t_c$, there is $\Delta P<0$, 
implying that the control is totally inefficient with these parameters.

Fig.\ \ref{FIG:PicD} shows the results with a qubit-oscillator detuning where $\omega_E=10\omega_S$. Other parameters are the
same as those in Fig.\ \ref{FIG:PicC}. Typically, the parity kicks are inefficient if $t_c$ is large as seen in
Fig.\ \ref{FIG:PicD}(a) and (b). Moreover, the strengths of memory effects are weaker compared with those in 
Fig.\ \ref{FIG:PicC} as seen in Fig.\ \ref{FIG:PicD}(c) and (d).  Similar to the results in Fig.\ \ref{FIG:PicB}, 
the relations $\tilde{N}_M^{t_{1:2}}\approx\bar{N}_M^{t_{1:2}}\approx2E$ and  
 $\tilde{N}_M^{t_{1:2}}[\rho_S(0)]\approx\bar{N}_M^{t_{1:2}}[\rho_S(0)]\approx2E[\rho_S(0)]$ approximately hold 
for a wide range of $t_c$ in Fig.\ \ref{FIG:PicD}(c) and (d). 

Additionally, we provide in Fig.\ \ref{FIG:PicAHeat} the results with the same parameters as in Fig.\ \ref{FIG:PicA} 
but $\bar{n}=5$ to investigate the influence of temperature on the role of memory effects in one cycle of parity kicks.
It is observed that the dynamics is significantly altered by the increased temperature of the environment and the
 performances with control decay faster compared with those in Fig.\ \ref{FIG:PicA}. 
Remarkably, in Fig.\ \ref{FIG:PicAHeat}(c) and (d), the upper bounds and the lower bounds become closer compared with Fig.\ \ref{FIG:PicA}. 
The reason is that the strength of memory effects $\bar{N}_M^{t_{1:2}}$  ($\bar{N}_M^{t_{1:2}}[\rho_S(0)]$) is suppressed more 
significantly by the increased temperature compared with $\tilde{N}_M^{t_{1:2}}$ ($\tilde{N}_M^{t_{1:2}}[\rho_S(0)]$).
Similar to the results in Fig.\ \ref{FIG:PicA}(c) and (d), $\Delta P\approx E$ and $\Delta P[\rho_S(0)]\approx E[\rho_S(0)]$
are satisfied for a wide range of $t_c$. Specially, there are  $\Delta P\approx E\approx \tilde{N}_M^{t_{1:2}}$ and 
$\Delta P[\rho_S(0)]\approx E[\rho_S(0)]\approx \tilde{N}_M^{t_{1:2}}[\rho_S(0)]$ when $t_f\approx5.7/g$.

In Fig.\ \ref{FIG:PicA}-\ref{FIG:PicAHeat} and other further simulations with one cycle of parity kicks,  we find that
the following properties hold in general (at least) when $t_c$ is small enough. (1) The effects of control are close to their upper bounds.
(2) The increases of performances are close to the effects of control. (3) The lower bounds of memory effects are close to zero. 
Therefore, there are $\tilde{N}_M^{t_{1:2}}\approx\bar{N}_M^{t_{1:2}}\approx2E\approx2\Delta P$ and
 $\tilde{N}_M^{t_{1:2}}[\rho_S(0)]\approx\bar{N}_M^{t_{1:2}}[\rho_S(0)]\approx2E[\rho_S(0)]\approx2\Delta P[\rho_S(0)]$.
 The above properties imply that the increase of the performances (by the control) or the effects of control  
can directly reflect the strengths of memory effects under certain conditions. 

\subsection{Results for many cycles of parity kicks}

 \begin{figure}
\includegraphics*[width=9.2cm]{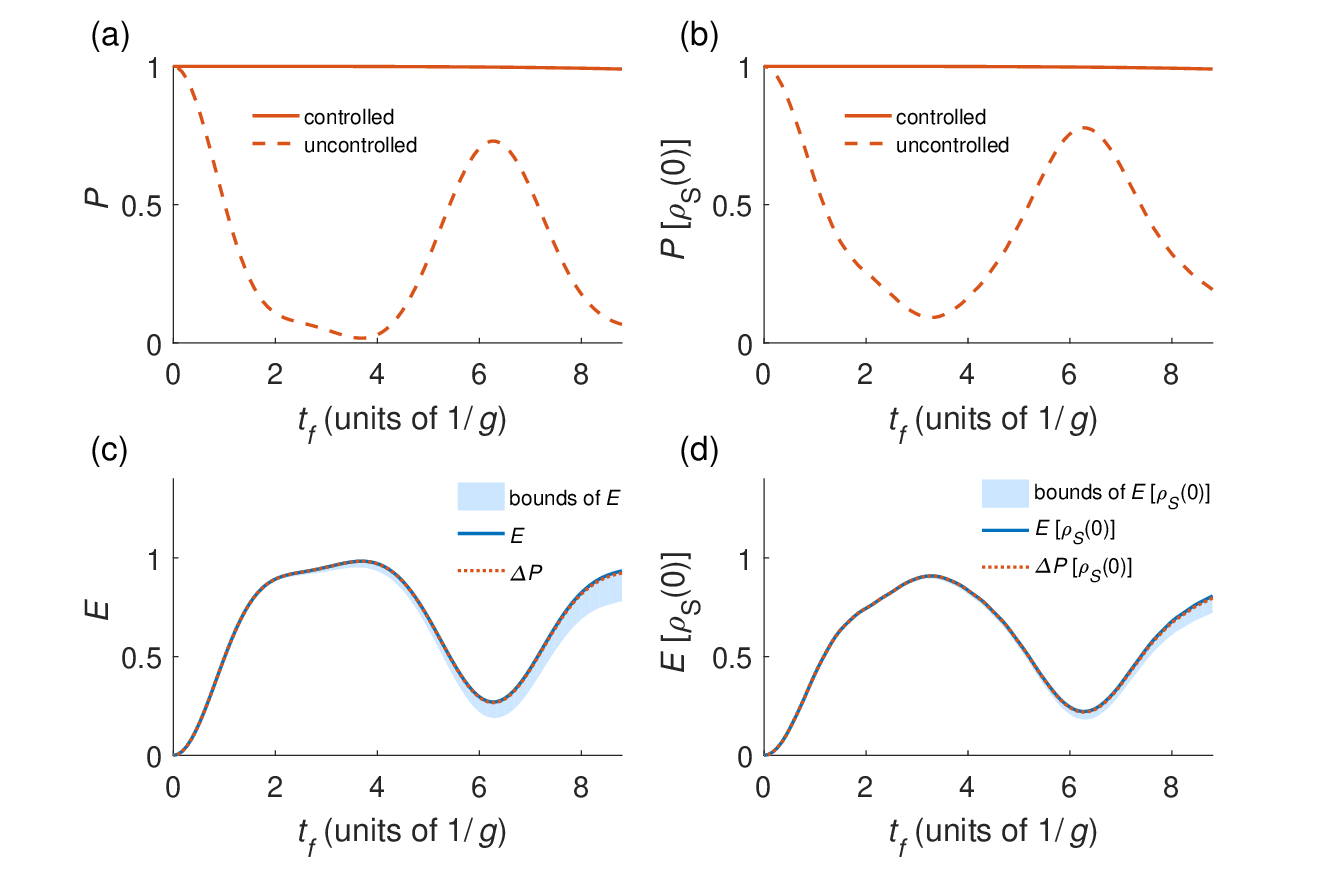}
\caption{Performances (with and without control), effects of control and their bounds as a function of the final time $t_f$ 
for 500 cycle of parity kicks. Other parameters and settings are the same as those in Fig.\ \ref{FIG:PicA}.  }
\label{FIG:PicA500}
\end{figure}

The role of memory effects in more than one cycle of parity kicks ($k\geqslant2$) is discussed in the following where  
$\tilde{N}_M^{t_{1:n}}$, $\bar{N}_M^{t_{1:n}}$, $\tilde{N}_M^{t_{1:n}}[\rho_S(0)]$ and $\bar{N}_M^{t_{1:n}}[\rho_S(0)]$
are calculated with $t_f=kt_c=n\tau$ ($n=2k\geqslant4$). For simplicity, we will consider scenarios with a large number of cycles 
and a finite final time $t_f$ so that the control is expected to be efficient. Fig.\ \ref{FIG:PicA500} shows the result 
with $500$ cycles of parity kicks ($t_f=500t_c, n=1000$) where other parameters and settings are the same as those 
in Fig.\ \ref{FIG:PicA}. It is observed in Fig.\ \ref{FIG:PicA500}(a) and (b) that the performances $\Delta P$ and 
 $\Delta P[\rho_S(0)]$ under control almost reach their maxima due to the frequent kicks. Meanwhile, the gap 
 between $E^{\mathrm{ub}}$ and $E^{\mathrm{lb}}$ or $E^{\mathrm{ub}}[\rho_S(0)]$ and $E^{\mathrm{lb}}[\rho_S(0)]$
 is much smaller compared with that in Fig.\ \ref{FIG:PicA}(c) or (d), especially when $t_c$ is small. The reason is 
that when the number of cycles is large, the high-frequency kicks efficiently decouple the system from its 
environment so that the system dynamics is almost unitary and Markovian, i.e.,  $\tilde{N}_M^{t_{1:n}}\approx0$ 
and $\tilde{N}_M^{t_{1:n}}[\rho_S(0)]\approx0$. Correspondingly, there are 
$E^{\mathrm{ub}}\approx E^{\mathrm{lb}}\approx\bar{N}_M^{t_{1:n}}\approx E\approx \Delta P$ and
$E^{\mathrm{ub}}[\rho_S(0)]\approx E^{\mathrm{lb}}[\rho_S(0)]\approx\bar{N}_M^{t_{1:n}}[\rho_S(0)]\approx E[\rho_S(0)]\approx \Delta P[\rho_S(0)]$.
For sufficiently frequent kicks, these relations also hold for the regimes in Fig.\ \ref{FIG:PicB}-\ref{FIG:PicAHeat} as 
verified by our simulations (not shown for simplicity). We conclude that in the limit of infinite frequent kicks, 
the above relations holds exactly. That is, the memory effects without control fully account for the success of 
the dynamical decoupling task with infinite frequent kicks. In Fig.\ \ref{FIG:PicA500}(c) and (d), it is seen that
$\bar{N}_M^{t_{1:n}}$ or $\bar{N}_M^{t_{1:n}}[\rho_S(0)]$ may approach its maximum 1 in some cases. Similarly, in a 
near-Markovian regime like that in Fig.\ \ref{FIG:PicB}, $\bar{N}_M^{t_{1:n}}$ or $\bar{N}_M^{t_{1:n}}[\rho_S(0)]$ 
might also approach its maximum if the kicks are frequent enough, but not implying strong memory effects.
The reason is that longer $t_f$ and more initializations of environment (more kicks) are required to 
obtain a large $\bar{N}_M^{t_{1:n}}$ or $\bar{N}_M^{t_{1:n}}[\rho_S(0)]$ in a near-Markovian regime as discussed before.

\begin{figure}
\includegraphics*[width=9.2cm]{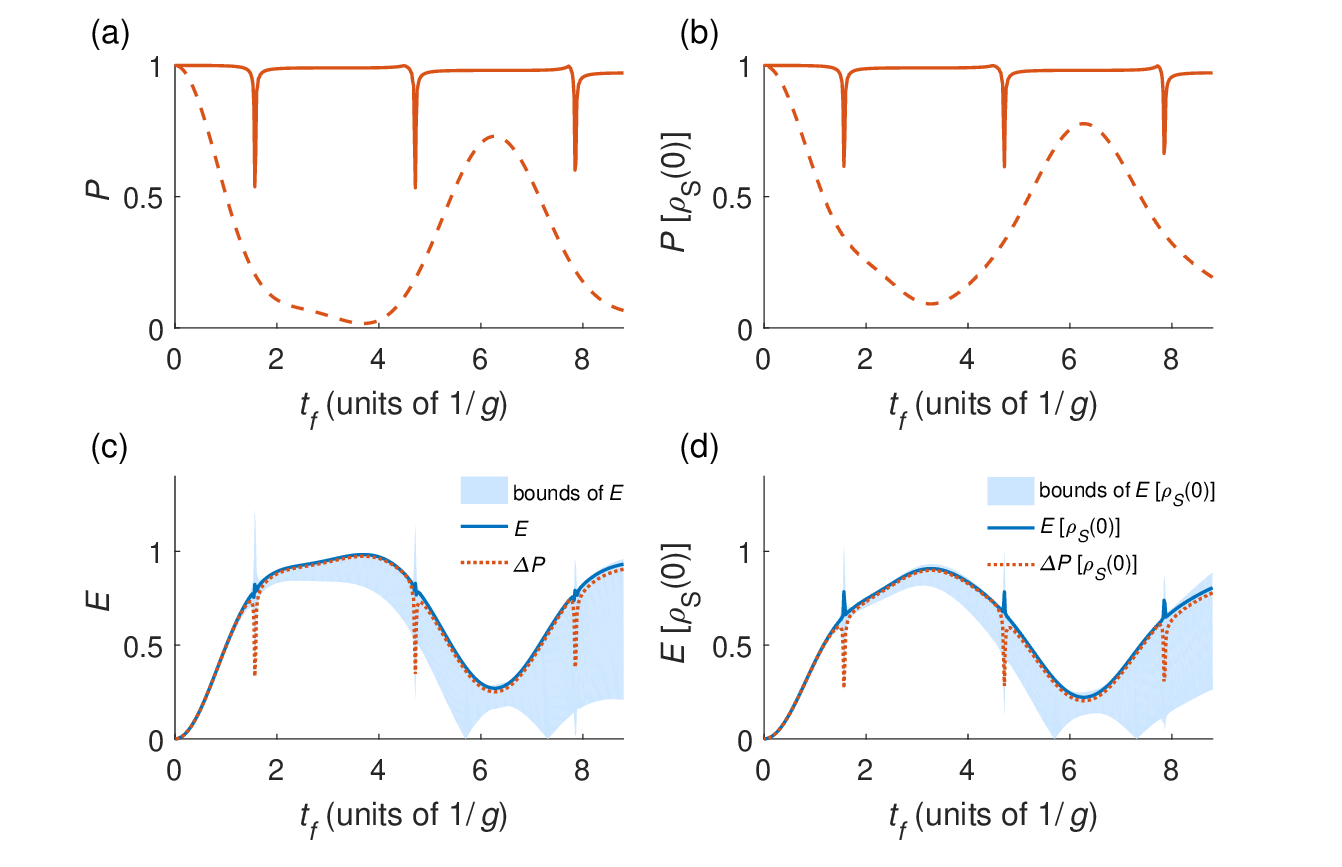}
\caption{Performances (with and without control), effects of control and their bounds as a function of the final time $t_f$ 
for 50 cycle of parity kicks. Other parameters and settings are the same as those in Fig.\ \ref{FIG:PicA500}. }
\label{FIG:PicA50}
\end{figure}

Interestingly, we find that when the period (or frequency) of the kicks reaches 
particular values, the control may become inefficient abruptly. For example, typically in a regime with $\omega_S=\omega_E$,
$g/\omega_S \lesssim 0.1$ and a weak oscillator dissipation as that in Fig.\ \ref{FIG:PicA}, when the cycle time $t_c$ 
satisfies $t_c=n_{\mathrm{odd}}t^{\mathrm{fast}}$ where $n_{\mathrm{odd}}=1,3,5\cdots$ and 
$t^{\mathrm{fast}}=2\pi/(\omega_S+\omega_E)$ is the period of the fast oscillations appeared in the counter-rotating 
terms in the interaction picture, the performance with parity kicks may decrease significantly. An example is shown 
in Fig.\ \ref{FIG:PicA50} where $k=50$ and other parameters (settings) are the same as those in Fig.\ \ref{FIG:PicA500}.
For the parameters in Fig.\ \ref{FIG:PicA50}, the period of the fast oscillation is $t^{\mathrm{fast}}=\pi/(100g)$ and $t_c=t_f/50$. 
It is seen in Fig.\ \ref{FIG:PicA50}(a) and (b) that when $t_c=\pi/(100g),3\pi/(100g),5\pi/(100g)$, the performances 
of parity kicks drop significantly. Further simulations shows that this phenomenon does not occur if the qubit-oscillator 
coupling is of the JC type. Therefore, it could be physically explained by the theory in \cite{Huang} as the enhancement 
of counter-rotating interactions by the parity kicks with a proper frequency. Besides, as seen in Fig.\ \ref{FIG:PicA50}(c) 
and (d), the gap between $E^{\mathrm{ub}}$ and $E^{\mathrm{lb}}$ or $E^{\mathrm{ub}}[\rho_S(0)]$ and $E^{\mathrm{lb}}[\rho_S(0)]$
 is smaller than that in Fig.\ \ref{FIG:PicA500}(a) or (b) due to the lower frequency of kicks.


\section{Conclusion and discussion}
In this work, we investigate quantitative connections between the environmental memory effects and 
the characteristics of a dynamical decoupling process, especially, that controlled by parity kicks. 
Considering that there exists two special time instants (corresponding to two kicks) in one cycle of parity 
kicks which are linked to the past-future dependence of the system, we tailor our measurements of memory 
effects to evaluate the strengths of memory effects associated with two specific time instants. For a 
decoupling process with more than 2 specific time instants (e.g., many cycles of parity kicks), we 
extend our measurements of memory effects to their multi-time forms. Our measures of memory effects in 
terms of dynamical maps and quantum states can be interpreted as the past-future dependence of the system 
dynamics and is easy to calculate. Our main finding is that if the dynamical map of each kick commutes 
with the system's dynamical map without control, then the effect of dynamical decoupling 
($E$ or $E[\rho_S(0)]$) is upper (lower) bounded by the summation (difference) of the strengths 
of memory effects with and without control. That means (under certain conditions), 
the environmental memory effects with and without control together determine to what degree the system can 
be changed, and potentially, the performance (without control) can be increased by the control. 
On the other hand, the effect of the control or the increase of the performances brought by 
the control is a manifestation of non-Markovianity.
 
We propose a class of Hamiltonians and environment initial states that support the commutation 
relation for parity kicks. After that, we apply our theory to a dissipative quantum Rabi model where the 
qubit is controlled by parity kicks and the oscillator is in a Markovian reservoir at finite temperature. 
The effects of control, their bounds and the performances with and without control are numerically simulated 
and compared in different regimes, which verifies our finding.  Besides, the relationship among these quantities
 may take simple forms in particular regimes. For example, in one cycle of parity kicks, there is 
  $\tilde{N}_M^{t_{1:2}}\approx\bar{N}_M^{t_{1:2}}\approx2E\approx2\Delta P$ if the cycle time is short enough. 
 In the case of many cycles, if the frequency of kicks is high enough, there are 
$\bar{N}_M^{t_{1:n}}\approx E\approx \Delta P$ and $\tilde{N}_M^{t_{1:n}}\approx0$. Besides, it is found 
that the control might be inefficient with particular kick frequencies.  Our work provides new 
insights into the role of memory effects in dynamical decoupling and methods to measure the strength of 
memory effects by parity kicks.  

In our model, the Lindblad dissipators in Eq. (\ref{Eqn:ME}) and  Eq. (\ref{Eqn:MEcontrol}) are
phenomenological. Microscopically, a standard master equation is derived under the Born-Markov approximation implying 
a small $\gamma$ compared with $\omega_E$ \cite{Breuerbook}. Besides, the master equations (\ref{Eqn:ME}) 
and (\ref{Eqn:MEcontrol}) imply that the dissipation of the harmonic oscillator is 
independent of the qubit, which is valid when the qubit-oscillator 
coupling $g$ is not strong \cite{Beaudoin} compared with $\omega_S$, $\omega_{E}$. Therefore, a standard 
master equation typically can not accurately described the dynamics for large $\gamma$ or $g$. However, 
our model provides a phenomenological description with highly adjustable non-Markovianity and completely
positive dynamics. Besides, such model (and related models) severs as a building block to simulate 
complex environments as used in the pseudomode theory. Experimentally, Eq.\ (\ref{Eqn:ME}) at zero temperature 
might be effectively achieved by a driven trapped ion system \cite{Hwang2018}.

\section*{ACKNOWLEDGMENTS}
The authors thank S. Xu for helpful discussions.  This work is supported by
the National Natural Science Foundation of China under Grant No. 11705026 and No. 12005033, 
the Fundamental Research Funds for the Central Universities in China under 
Grant No. 3132020178 and No. 3132025192.

\appendix

\section{DERIVATIONS OF THE BOUND CONDITIONS EQ.\ (\ref{Eqn:PKvsNM}) AND (\ref{Eqn:PKvsNMstate})} 
\label{Sec:AppendixA}

In this appendix, we present the derivation of the bound conditions Eq.\ (\ref{Eqn:PKvsNM}) and (\ref{Eqn:PKvsNMstate})
 for one cycle of parity kicks. Remind that the strength of memory effects with and without one cycle of parity kicks are defined as  
\begin{eqnarray}
\tilde{N}_M^{t_{1:2}}&=&D[\rho_{\tilde{T}(2\tau,0)},\rho_{\tilde{T}(2\tau,\tau)\tilde{T}(\tau,0)}],\\
\label{EqnA:NMt1t2PK0}
\bar{N}_M^{t_{1:2}} &=& D[\rho_{\bar{T}(2\tau,0)},\rho_{\bar{T}(2\tau,\tau)\bar{T}(\tau,0)}]
\label{EqnA:NMt1t2PK}
\end{eqnarray}
in Eq.\ (\ref{Eqn:NMt1t2PK}) and (\ref{Eqn:NMt1t2PK0}), respectively. 
Within Eq.\ (\ref{Eqn:TcKT0}), it is observed that if $\mathcal{K}_P\bar{T}(t_b,t_a)=\bar{T}(t_b,t_a)\mathcal{K}_P$, 
there is
\begin{eqnarray}
\tilde{T}(2\tau,\tau)\tilde{T}(\tau,0)&=&\mathcal{K}_PT(2\tau,\tau)\mathcal{K}_PT(\tau,0)\nonumber\\
&=&\mathcal{K}_P\mathcal{K}_P\bar{T}(2\tau,\tau)\bar{T}(\tau,0)\nonumber\\
&=&\bar{T}(2\tau,\tau)\bar{T}(\tau,0)
 \label{EqnA:KTKT}
\end{eqnarray}
where $\mathcal{K}_P\mathcal{K}_P=\mathbb{I}$ is used due to the properties of parity operators 
or Eq.\ (\ref{Eqn:condition}). Accordingly, the Choi-Jami\'{o}{\l}kowski matrices of the left-hand side and
right-hand side of Eq.\ (\ref{EqnA:KTKT}) satisfy
\begin{eqnarray}
\rho_{\tilde{T}(2\tau,\tau)\tilde{T}(\tau,0)}=\rho_{\bar{T}(2\tau,\tau)\bar{T}(\tau,0)}.
 \label{EqnA:cTcTTT}
\end{eqnarray}
Note that the trace distance obeys the triangle inequality 
\begin{eqnarray}
D(\rho_1,\rho_2)\leqslant D(\rho_1,\rho_3) +D(\rho_2,\rho_3) 
\label{EqnA:triangle}
\end{eqnarray}
and its variant 
\begin{eqnarray}
|D(\rho_1,\rho_3)-D(\rho_2,\rho_3)|\leqslant D(\rho_1,\rho_2).
\label{EqnA:triangle2}
\end{eqnarray}
Now let $\rho_1=\rho_{\tilde{T}(2\tau,0)}$ and $\rho_2=\rho_{\bar{T}(2\tau,0)}$ so that    
 $E=D[\rho_1,\rho_2]=D[\rho_{\tilde{T}(2\tau,0)},\rho_{\bar{T}(2\tau,0)}]$ by the definition of the 
effect of two parity kicks. If $\mathcal{K}_P$ commutes with $\bar{T}(t_b,t_a)$, we can set $\rho_3$ 
to be $\rho_{\tilde{T}(2\tau,\tau)\tilde{T}(\tau,0)}$ or $\rho_{\bar{T}(2\tau,\tau)\bar{T}(\tau,0)}$
according to Eq.\ (\ref{EqnA:cTcTTT}). Specifically, let one of $\rho_3$ in Eq.\ (\ref{EqnA:triangle}) 
or Eq.\ (\ref{EqnA:triangle2}) be $\rho_{\tilde{T}(2\tau,\tau)\tilde{T}(\tau,0)}$ and the other 
 be $\rho_{\bar{T}(2\tau,\tau)\bar{T}(\tau,0)}$, then Eq.\ (\ref{EqnA:triangle}) and 
Eq.\ (\ref{EqnA:triangle2}) together can be expressed as
\begin{eqnarray}
 |\tilde{N}_M^{t_{1:2}}-\bar{N}_M^{t_{1:2}}|\leqslant  E \leqslant \tilde{N}_M^{t_{1:2}}+\bar{N}_M^{t_{1:2}}
 \label{EqnA:PKvsNM}
\end{eqnarray}
according to the definitions of $\tilde{N}_M^{t_{1:2}}$ and $\bar{N}_M^{t_{1:2}}$.

For one cycle of parity kicks, the bound condition Eq.\ (\ref{Eqn:PKvsNMstate}) in terms of 
quantum states can be derived similarly. By acting $\tilde{T}(2\tau,0)$, $\tilde{T}(2\tau,\tau)\tilde{T}(\tau,0)$, 
$\bar{T}(2\tau,0)$  and $\bar{T}(2\tau,\tau)\bar{T}(\tau,0)$ on the initial state $\rho_S(0)$, we get 4 final states  
 $\tilde{\rho}_S(t_c)$, $\tilde{\rho}'_S(t_c)$, $\bar{\rho}_S(t_c)$ and $\bar{\rho}'_S(t_c)$, respectively. 
 If $\mathcal{K}_P$ commutes with $\bar{T}(t_b,t_a)$, there is
 \begin{eqnarray}
\tilde{\rho}'_S(t_c)=\bar{\rho}'_S(t_c).
 \label{EqnA:crhoprhop}
\end{eqnarray}
according to Eq.\ (\ref{EqnA:KTKT}). In Eq.\ (\ref{EqnA:PKvsNM}), by using the final states 
 $\tilde{\rho}_S(t_c)$, $\tilde{\rho}'_S(t_c)$, $\bar{\rho}_S(t_c)$ and $\bar{\rho}'_S(t_c)$ instead 
 of $\rho_{\tilde{T}(2\tau,0)}$, $\rho_{\tilde{T}(2\tau,\tau)\tilde{T}(\tau,0)}$, $\rho_{\bar{T}(2\tau,0)}$
  and $\rho_{\bar{T}(2\tau,\tau)\bar{T}(\tau,0)}$, respectively, we get the following condition 
\begin{eqnarray}
 &&|\tilde{N}_M^{t_{1:2}}[\rho_S(0)]-\bar{N}_M^{t_{1:2}}[\rho_S(0)]|\leqslant E[\rho_S(0)] \nonumber\\ 
 &&\leqslant \tilde{N}_M^{t_{1:2}}[\rho_S(0)]+\bar{N}_M^{t_{1:2}}[\rho_S(0)]
 \label{EqnA:PKvsNMstate}
\end{eqnarray}
according to the definitions (\ref{Eqn:NMIt1t2PK}), (\ref{Eqn:NMIt1t20}) and $E[\rho_S(0)] = D[\tilde{\rho}_S(t_c),\bar{\rho}_S(t_c)]$.

 \section{DERIVATIONS OF THE BOUND CONDITIONS EQ.\ (\ref{Eqn:DDvsNM}) AND (\ref{Eqn:DDvsNMstate})} 
 \label{Sec:AppendixB}
In the case of a general dynamical decoupling process with $n$  kicks ($n\geqslant2$) and $t_f=n\tau=kt_c$, 
 the bound conditions Eq.\ (\ref{Eqn:DDvsNM}) and (\ref{Eqn:DDvsNMstate}) can be derived similarly. 
 Within Eq.\ (\ref{Eqn:TcKT0}), if each kick $\mathcal{K}_j$ commutes with $\bar{T}(t_b,t_a)$, there is  
\begin{eqnarray}
&&\tilde{T}[n\tau,(n\!-\!1)\tau]\cdots\tilde{T}(2\tau,\tau)\tilde{T}(\tau,0) \nonumber\\
&=&\mathcal{K}_n\bar{T}[n\tau,(\!n\!-\!1\!)\tau]\cdots \mathcal{K}_2\bar{T}(2\tau,\tau)\mathcal{K}_1\bar{T}(\tau,0)\nonumber\\
&=&\mathcal{K}_n \cdots\mathcal{K}_2\mathcal{K}_1\bar{T}[n\tau,(\!n\!-\!1\!)\tau]\cdots\bar{T}(2\tau,\tau)\bar{T}(\tau,0)\quad\nonumber\\
&=&\bar{T}[n\tau,(\!n\!-\!1\!)\tau]\cdots\bar{T}(2\tau,\tau)\bar{T}(\tau,0)
\label{EqnA:GKTKT}
\end{eqnarray}
where the properties Eq.\ (\ref{Eqn:condition}) is used. It follows that  
 \begin{eqnarray}
&&\rho_{\tilde{T}[n\tau,(n\!-\!1)\tau]\cdots\tilde{T}(2\tau,\tau)\tilde{T}(\tau,0)}\nonumber\\
&=&\rho_{\bar{T}[n\tau,(\!n\!-\!1\!)\tau]\cdots\bar{T}(2\tau,\tau)\bar{T}(\tau,0)}.
 \label{EqnA:GcTcTTT}
\end{eqnarray}
By using $\rho_{\tilde{T}(n\tau,0)}$, $\rho_{\tilde{T}[n\tau,(n\!-\!1)\tau]\cdots\tilde{T}(2\tau,\tau)\tilde{T}(\tau,0)}$, 
$\rho_{\bar{T}(n\tau,0)}$ and  $\rho_{\bar{T}[n\tau,(\!n\!-\!1\!)\tau]\cdots\!\bar{T}(2\tau,\tau)\bar{T}(\tau,0)}$ 
instead of $\rho_{\tilde{T}(2\tau,0)}$, $\rho_{\tilde{T}(2\tau,\tau)\tilde{T}(\tau,0)}$, $\rho_{\bar{T}(2\tau,0)}$ 
and  $\rho_{\bar{T}(2\tau,\tau)\bar{T}(\tau,0)}$, respectively in Eq.\ (\ref{EqnA:PKvsNM}), we obtain  
\begin{eqnarray}
 |\tilde{N}_M^{t_{1:n}}-\bar{N}_M^{t_{1:n}}|\leqslant  E \leqslant \tilde{N}_M^{t_{1:n}}+\bar{N}_M^{t_{1:n}}
 \label{EqnA:DDvsNM}
\end{eqnarray}
according to the definitions (\ref{Eqn:NMt1tnDD}), (\ref{Eqn:NMt1tn0}) and $E=D[\rho_{\tilde{T}(n\tau,0)},\rho_{\bar{T}(n\tau,0)}]$.

Let $\tilde{T}(n\tau,0)$, $\tilde{T}[n\tau,(n\!-\!1)\tau]\cdots\tilde{T}(2\tau,\tau)\tilde{T}(\tau,0)$, 
$\bar{T}(n\tau,0)$  and $\bar{T}[n\tau,(\!n\!-\!1\!)\tau]\cdots\!\bar{T}(2\tau,\tau)\bar{T}(\tau,0)$ act on 
the initial state $\rho_S(0)$, we get 4 final states  $\tilde{\rho}_S(kt_c)$, $\tilde{\rho}'_S(kt_c)$, 
$\bar{\rho}_S(kt_c)$ and $\bar{\rho}'_S(kt_c)$, respectively. If each kick $\mathcal{K}_j$ 
commutes with $\bar{T}(t_b,t_a)$, there is 
 \begin{eqnarray}
\tilde{\rho}'_S(kt_c)=\bar{\rho}'_S(kt_c).
 \label{EqnA:Gcrhoprhop}
\end{eqnarray}
according to Eq.\ (\ref{EqnA:GKTKT}). Then using $\tilde{\rho}_S(kt_c)$, $\tilde{\rho}'_S(kt_c)$,
 $\bar{\rho}_S(kt_c)$ and $\bar{\rho}'_S(kt_c)$ instead of $\rho_{\tilde{T}(n\tau,0)}$, 
 $\rho_{\tilde{T}[n\tau,(n\!-\!1)\tau]\cdots\tilde{T}(2\tau,\tau)\tilde{T}(\tau,0)}$, 
 $\rho_{\bar{T}(n\tau,0)}$ and $\rho_{\bar{T}[n\tau,(\!n\!-\!1\!)\tau]\cdots\!\bar{T}(2\tau,\tau)\bar{T}(\tau,0)}$, 
 respectively in Eq.\ (\ref{EqnA:DDvsNM}), there is
\begin{eqnarray}
 &&|\tilde{N}_M^{t_{1:n}}[\rho_S(0)]-\bar{N}_M^{t_{1:n}}[\rho_S(0)]|\leqslant E[\rho_S(0)] \nonumber\\ 
 &&\leqslant \tilde{N}_M^{t_{1:n}}[\rho_S(0)]+\bar{N}_M^{t_{1:n}}[\rho_S(0)]
 \label{EqnA:DDvsNMstate}
\end{eqnarray}
according to the definitions \ (\ref{Eqn:NMt1tnDD}), (\ref{Eqn:NMt1tn0}) and $E[\rho_S(0)] = D[\tilde{\rho}_S(kt_c),\bar{\rho}_S(kt_c)]$.

\end{document}